\documentclass[twocolumn, superscriptaddress]{revtex4-2}

\usepackage{amsfonts}
\usepackage{amsmath}
\usepackage{amsthm}
\usepackage{graphicx}
\usepackage{xcolor}
\usepackage{stmaryrd}
\usepackage{dsfont}
\usepackage{comment}
\usepackage{multirow}
\usepackage{relsize}

\usepackage[colorlinks=true, allcolors=blue]{hyperref}
\usepackage[capitalise]{cleveref}
\usepackage{orcidlink}

\newcommand{\ex}[1]{\textsc{#1 example:} }

\begin{document}

\title{A matching decoder for bivariate bicycle codes}

\author{Kaavya Sahay}

\affiliation{Yale University, Department of Applied Physics, New Haven, Connecticut 06520, USA}

\affiliation{Yale Quantum Institute, Yale University, New Haven, Connecticut 06511, USA}

\affiliation{IBM Quantum, T. J. Watson Research Center, Yorktown Heights, New York 10598, USA}

\author{Dominic J. Williamson}
\thanks{Current address: School of Physics, University of Sydney, Sydney, NSW 2006, Australia}

\affiliation{IBM Quantum, IBM Almaden Research Center, San Jose, CA 95120, USA}

\author{Benjamin J. Brown}
\affiliation{IBM Quantum, T. J. Watson Research Center, Yorktown Heights, New York 10598, USA}

\affiliation{IBM Denmark, Sundkrogsgade 11, 2100 Copenhagen, Denmark}

\begin{abstract}
The discovery of new quantum error-correcting codes that encode several logical qubits into relatively few physical qubits motivates the development of efficient and accurate methods of decoding these systems. Here, we adopt the minimum-weight perfect matching algorithm, a subroutine invaluable to decoding topological codes, to decode bivariate bicycle codes. 
Using the equivalence of bivariate bicycle codes to copies of the toric code, we propose a method we call the `cylinder trick' to rapidly find a correction using matching on code symmetries. 
We benchmark our decoder on the gross code family, cyclic hypergraph-product codes, generalized toric codes, and recently proposed directional codes under code capacity and phenomenological noise models, demonstrating the general applicability of our protocol.
For a subset of these codes, we find that our decoder can be significantly improved by augmenting matching with strategies including belief propagation and `over-matching', thus achieving performance competitive with state-of-the-art approaches.

\end{abstract}

\maketitle

\section{Introduction}

To protect logical  quantum information from physical errors, quantum-computing architectures require the redundant encoding of information into quantum error-correcting codes (QECCs)~\cite{Fowler2012surface, Lee2025low, Yoder2025tour}. QECCs rely on a \textit{decoder}~\cite{Dennis02, Duclos-Cianci2010fast, Wootton2012high, Bravyi2013quantum, Bravyi2014efficient, Anwar2014fast, Torlai2017neural, Delfosse2021almostlineartime, Brown2022conservation} to correct errors. A decoder is a classical algorithm that interprets syndrome data returned from a QECC to find a correction that reverses the effect of the incident error. 
Ideally, practical decoders minimize the logical error rate while also running in a time-efficient manner. \textit{Matching} decoders~\cite{Dennis02, Higgott_2025sparse, Wang2003confinement, Sahay2022decoder, Brown2022conservation} have been shown to perform well under these criteria for topological codes. 

Topological codes such as the toric code~\cite{KITAEV2003, Dennis02}, are QECCs supported on a lattice of physical qubits with geometrically local stabilizer operators.
Increasingly, quantum-computing architectures rely on non-local low-density parity check (LDPC) QECCs to encode a large amount of quantum information into relatively fewer physical qubits~\cite{Breuckmann2021quantum}. With the discovery of new LDPC codes, we require the development of decoding algorithms to support their operation~\cite{Panteleev2021degeneratequantum,Roffe_2020,müller2025improved,Hillmann2025localized,Wolanski2025ambiguity, ye2025beamsearchdecoderquantum}.

Here, we generalize matching decoders to bivariate bicycle (BB) codes~\cite{Bravyi2024high, Panteleev2021degeneratequantum,Kovalev2012Quantum}. These codes are defined on a two-dimensional lattice by a set of constant-sized stabilizer generators that are translations of one another over the lattice. 
To find a decoder, we use the equivalence between stabilizer codes defined on two-dimensional lattices and copies of the toric code~\cite{Chen2010local, Bombin2012universal, Bombin2014structure, haah2016algebraic, Haah2021classification,Chen2025anyon}. This equivalence means that any BB code shares an underlying global structure with copies of the toric code. 
We formalize this structure with the notion of a code \textit{symmetry}~\cite{Brown2020parallelized, Brown2022conservation}, which we exploit to find a correction using matching. 
Specifically, we introduce the \textit{cylinder trick}~\footnote{We note that Appendix~A of Ref.~\cite{Sahay2022decoder} proposes the cylinder trick for the explicit case of the color code. In this work we define the cylinder trick more generally}, which enables us to interpret edges returned from a matching subroutine in terms of the number of errors supported on a generating set of BB code logical operators. 
Given these numbers, it is straightforward to find a correction that is consistent with the result from a matching graph.

\begin{figure*}[t]
    \centering
    \includegraphics[width=1\linewidth]{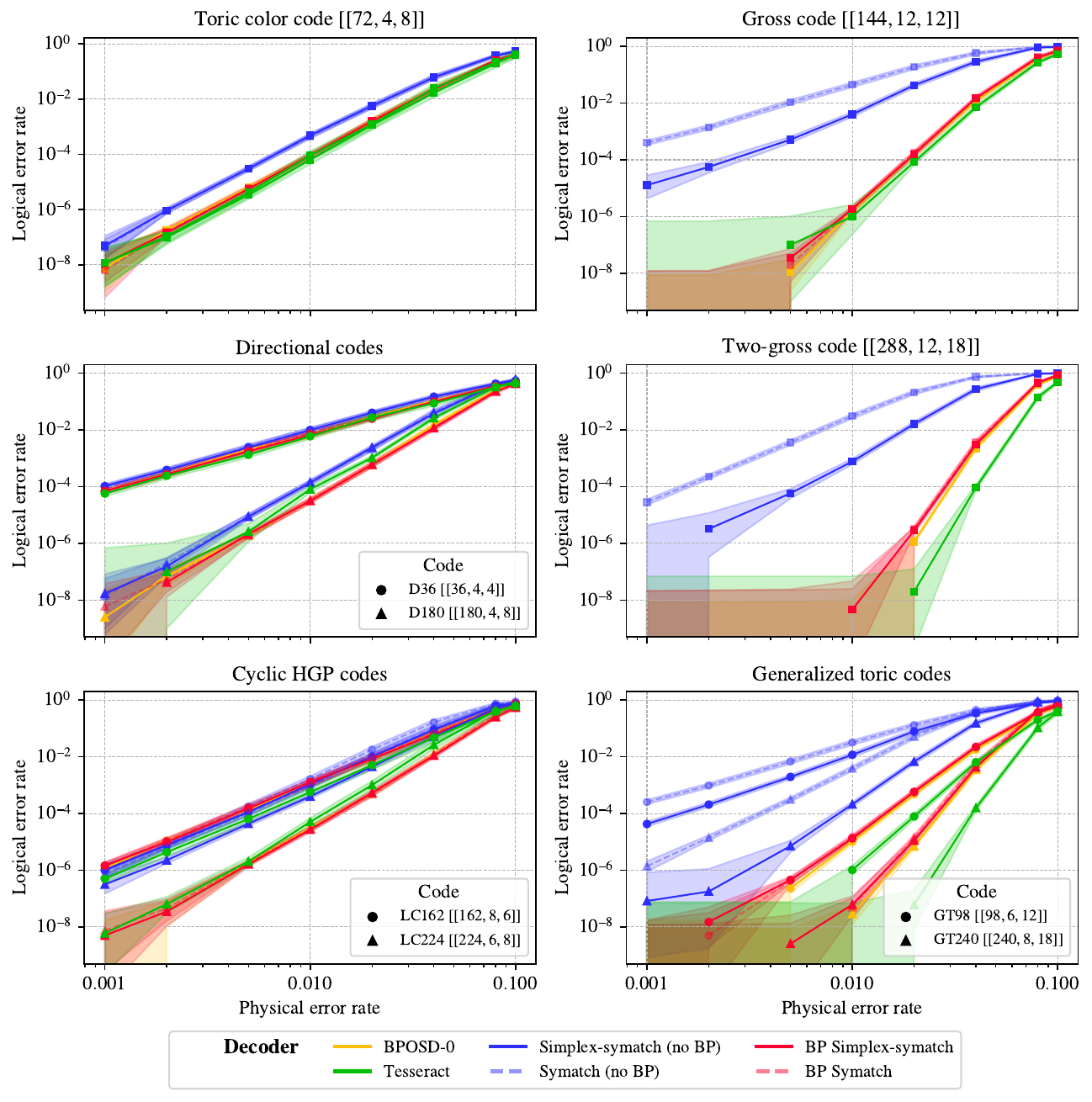}
    \caption{Evaluating decoder performance. The panels show  logical error rates versus physical error rates for the bivariate bicycle codes of \cref{table:codes} under code-capacity bit-flip noise for different decoders. We develop matching-based `symatch' and `simplex-symatch' decoders (blue and red). For codes on the left, our base strategies in blue tend to perform well. For codes on the right, bare symatch and simplex-symatch underperform; we find that augmentation by belief-propagation (BP) significantly improves logical error rates, as shown by the red curves. For all codes, our best decoders are competitive with state-of-the-art methods (green and yellow). All BP subroutines use the \texttt{min-sum} method for a maximum of 1000 iterations.}
    \label{fig:codecappl}
\end{figure*}

We benchmark the performance of our decoder, termed \textit{symatch}, on the gross code family~\cite{Bravyi2024high}, generalized toric codes~\cite{Liang_2025,Pecorari_2025high,aydin2025cyclichypergraphproductcode}, and directional codes~\cite{geher2025directional} under the code-capacity and phenomenological noise models. 
\cref{fig:codecappl} displays our code-capacity results. 
For a subset of these codes, our base decoder is distance preserving, demonstrating the first successful matching-based correction strategy for BB codes.
However, for specific instances, we find that bare matching can fail for errors of weight $w\ll d/2$, where $d$ is the code distance. We attribute this to the fact that stabilizers of practical BB codes, such as the gross code, have geometrically large support on the lattice. This means that a matching subroutine may choose low-weight incorrect paths. Additionally, matching on a symmetry ignores some syndrome information.

We therefore introduce new subroutines to improve the logical failure rate of our matching decoder. We call the first of these \textit{over-matching}. 
As we show, we can use consistency conditions between an overcomplete set of code symmetries to identify incorrect matching results. 
Second, we consider the use of efficient classical decoders to correct certain types of heralded low-weight error configurations.  
Finally,  we augment our matching subroutines with \textit{belief propagation}, i.e., use belief-matching~\cite{Higgott2023improved}. We find that this improves logical failure rates by furnishing matching subroutines with syndrome information not considered within a symmetry~\footnote{A similar idea has been used in the context of fracton codes in Ref.~\cite{SchwartzmanNowik2025generalizing}}.
The use of these additional subroutines enables correction of a significant majority of the failures due to low-weight errors. Consequently, our best decoder variants are competitive with current state-of-the art methods~\cite{Roffe_2020,beni2025tesseractsearchbaseddecoderquantum}, with reduced time complexity.

The remainder of this manuscript is organized as follows: In \cref{Sec:Prelim}, we provide definitions for BB codes and symmetries before describing our decoder in \cref{Sec:Decoder}. We lay out numerical results in \cref{Sec:Num}, and conclude in \cref{Sec:Conc}.

\section{Preliminaries}\label{Sec:Prelim}

In this Section we review concepts in the literature and introduce notation needed to describe our results. In~\cref{SubSec:BBcodes} we describe bivariate bicycle codes, and in~\cref{SubSec:CodeSymmetries,SubSec:Subsymmetries} we describe code symmetries essential to our decoder. In~\cref{SubSec:LocalEquivalence}, we review known results about local constant-depth circuit equivalence between QECCs defined on a two-dimensional lattice.

\subsection{Bivariate bicycle codes}
\label{SubSec:BBcodes}

Bivariate bicycle (BB) codes~\cite{Bravyi2024high} are LDPC QECCs defined on a two-dimensional translationally invariant (2DTI) lattice. BB codes are Calderbank-Shor Steane (CSS) codes, i.e. their stabilizer generators are composed of either exclusively $X$ operators or $Z$ operators acting on qubits~\cite{Calderbank1996good, Steane1996error}.
For CSS codes, we can decompose general Pauli errors into the product of a Pauli-$X$ error and a Pauli-$Z$ error and deal with them separately.  
In the case of BB codes, the set of $X$-stabilizer generators is equivalent to that of $Z$-stabilizers, up to a coordinate transformation which we explain shortly. We can therefore focus solely on the Pauli-$Z$ stabilizers to correct bit-flip errors and assume that equivalent results will hold when we correct $Z$ errors using Pauli-$X$ stabilizers.

\begin{figure}
\includegraphics{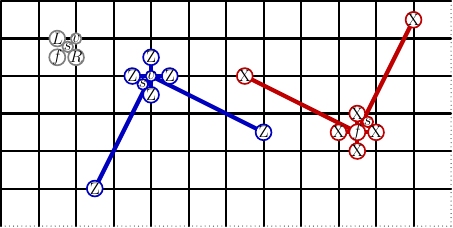}
\caption{\label{Fig:stabilizers} A square lattice with periodic boundary conditions, $(M,N, \alpha) = (12,6,0)$. (Top left) Each site $s$ includes a face $f$, a vertex $v$, and two qubits $L$ and $R$. For every site, qubit $L$ ($R$) lies on the horizontal (vertical) edge above (to the right of) the face of the respective site. (Center and right) A Pauli-$Z$ (Pauli-$X$) stabilizer for the gross code is shown in  blue (red), supported on a vertex $v$ (face $f$).}
\end{figure}

Here, we review BB codes using the polynomial formalism introduced in Ref.~\cite{Haah2013}. We also present the code in terms of check matrices in Appendix~\ref{App:CheckMatrix}.
In the polynomial formalism, monomials denote a translation about the two-dimensional lattice. For instance the term $x^jy^k$ denotes a translation of $j$ sites in the $x$ direction and $k$ sites in the $y$ direction. We can introduce boundary conditions by imposing constraints. Imposing $x^M = x^\alpha y^N = 1$ denotes a translationally invariant lattice with periodic boundary conditions of size $M \times N$, where $\alpha \neq 0$ indicates a twisted torus.

\begin{table*}[t]
\begin{center}
\begin{tabular}{|l||l|c|c|c|c|c|c|}
\hline
\textbf{Category} & \textbf{Name} & $\boldsymbol{[[n,k,d]]}$    & 
$\boldsymbol{M,N, \alpha}$ & $\boldsymbol{A}$ & $\boldsymbol{B}$ & $\boldsymbol{{kd^2}/{n}}$ \\
\hline 
\hline
\multirow{2}{*}{Topological codes~\cite{KITAEV2003}} & TC (toric code) & $[[2l^2,2,l]]$  &   $l,l, 0$ &  $1+x$ & $1+y$ & 1\\
& CC (toric color code) & $[[2l^2,4,4l/3]]$  &   $l,l, 0$ &  $1+x+y$ & $1+y+x^{-1}y$ & 3.56\\
\hline 
\multirow{2}{*}{Gross code family~\cite{Bravyi2024high}} & BB144 (gross code) & $[[144,12,12]]$ &   $12,6,0$ &  $1+x+x^{-1}y^{3}$ & $1+y+y^{-1}x^{3}$ & 12 \\
& BB288 (two-gross code) & $[[288,12,18]]$ &   $12,12,0$ & $1+x+x^{-1}y^{-3}$  & $1+y+y^{-1}x^{3}$ & 13.5\\
\hline
\multirow{2}{*}{Generalized toric codes~\cite{Liang_2025}} & GT98  & $[[98,6,12]]$ &   $7,7,0$ & $1+x+x^{-1}y^{-2}$  & $1+y+y^{-1}x^{2}$ & 8.82\\
& GT240  & $[[240,12,18]]$ &   $10,12,3$ & $1+x+x^{2}y$  & $1+y+y^{-2}x$ & 10.8\\
\hline
\multirow{2}{*}{$NE^3N$ directional codes~\cite{geher2025directional}} & D36 & $[[36,4,4]]$ &   $9,2,0$ & $1+x^3y^{-1}$  & $1+x+x^2$ & 1.78\\
& D180  & $[[180,4,8]]$ &   $15,9,6$ & $x^{-1}+x^2y^{-1}$  & $1+x+x^2$ & 1.42 \\
\hline
\multirow{2}{*}{Cyclic HGP codes~\cite{aydin2025cyclichypergraphproductcode}} & LC162 (La-cross~\cite{Pecorari_2025high}) & $[[162,8,6]]$ &   $9,9,0$ & $1+x+x^2$  & $1+y+y^2$ & 1.78\\
& LC224  & $[[224,6,8]]$ &   $14,8,0$ & $1+x+x^3$  & $1+y$ & 1.71 \\
\hline
\end{tabular} 
\end{center}
\caption{Examples of relevant BB codes and their parameters.  The codes studied have weight-4 to weight-6 checks and check matrices  $H^Z=[A|B]$
and $H^X=[B^T|A^T]$ with  $A$ and $B$ defined in polynomial form. 
}
\label{table:codes}
\end{table*}

We can write Pauli-$Z$ operators up to phases on sites of this translationally invariant lattice using a polynomial with variables $x$ and $y$ over the field $\mathbb{F}_2$:
\begin{equation}
q = \sum_{j,k} q_{j,k} x^j y^k,
\end{equation}
where coefficients $q_{j,k}$  take values 0 or 1. The vector $q$ concisely denotes the Pauli operator
\begin{equation}
P(q) = \prod_{j,k} Z_{(j,k)}^{q_{j,k}}.
\end{equation}

For BB codes, we introduce two qubits per site by extending $q$ into a $V$ column vector $ q = \left( q_1   ~~q_2 \right)^T $
where now $q_i$ are polynomials denoting the support of the Pauli-$Z$ operators on the $i$-th qubit at each site. In the literature, these are commonly known as left ($i=1$) and right ($i=2$) qubits. We will illustrate the code with qubits on the edges of a square lattice, where each site $s$ has its left qubit $L$ on a horizontal edge and its right qubit $R$ on a vertical edge; see Fig.~\ref{Fig:stabilizers} (top left).

This language gives us a compact way of writing the stabilizer group. 
We write a Pauli-$Z$ stabilizer generator for bivariate bicycle codes as
\begin{equation}
\sigma = \begin{pmatrix} A \\ B \end{pmatrix}.
\label{Eqn:Pauli-$Z$stabilizers}
\end{equation}
where $A$, $B$ are polynomials with either two or three terms. 
The support of the Pauli-$Z$ type stabilizer on the $L$ and $R$ qubits can be read off from the terms of the $A$ and $B$ polynomials respectively when they are interpreted as translations relative to the origin location, $1$.
We can now write an arbitrary element of the stabilizer group as $ q \sigma $
for some arbitrary choice of polynomial $q$. Indeed, the monomial terms in $q$ denote different translations of the stabilizer generators associated with vertices $v$ of the lattice.

We note we have an equivalent Pauli-$X$ type stabilizer module with the following transformation $A \rightarrow B^{-1}$ and $B\rightarrow A^{-1}$ where the inverse notation replaces all the terms in the polynomial with their inverse, i.e. translations in the negative direction. It follows that we require that polynomials $AB + B^T A^T = 0$ such that the Pauli-$X$ and Pauli-$Z$ stabilizers commute~\cite{Haah2013}.

A BB code can be completely defined by the choice of  polynomials $A$ and $B$ — which determine the stabilizer connectivity, and $(M,N,\alpha)$ — which determine the periodic boundary conditions~\footnote{We restrict our attention to codes with at most one twisted boundary.}. We list examples of BB codes defined by these parameters in \cref{table:codes}. We will refer to these codes by the names listed in the second column for the remainder of this work.

\ex{BB144} The gross code has $(M,N,\alpha) = (12,6,0)$ and polynomials
\begin{equation}
 A = 1 + x + x^{-1}y^{3}, \quad B = 1 + y + y^{-1}x^3.\label{Eqn:AandB}
\end{equation} 
Let us look at how a stabilizer can be read directly from the $\sigma$ of \cref{Eqn:Pauli-$Z$stabilizers}.  We will consider the Pauli-$Z$ stabilizer shown in the middle of Fig.~\ref{Fig:stabilizers}. This stabilizer has polynomial $A$ support on the horizontal (left) qubits and $B$ on the vertical (right) qubits of each site. The figure shows a Pauli-$Z$ operator on horizontal qubits at site $s$, a translation $x$ to the right, and another at a translation $x^{-1}y^{-3}$. 
Likewise, Pauli-$Z$ terms are supported on vertical qubits at translations $1$, $y$ and $x^3y^{-1}$ from the original site, i.e, the terms of polynomial $B$. All together, terms $A$ and $B$ denote the full support of the stabilizer on the left and right qubits.

\ex{Directional codes} Though not originally classified as BB codes in Ref.~\cite{geher2025directional}, this code family can readily be expressed using the polynomial formalism. We consider instances with weight-5 checks:
\begin{equation}
 A = 1 + x^{3}y^{-1}, \quad B = 1 + x + x^2.\label{Eqn:AandBDirn}
\end{equation} 

\ex{Cyclic HGP} Formalized in Ref.~\cite{aydin2025cyclichypergraphproductcode}, this hypergraph-product code family is a subclass of BB codes such that $A= A(x)$ and $B = B(y)$~\footnote{The La-cross code of Ref.~\cite{Pecorari_2025high} is a planar variant with redundant checks removed.}.

\begin{figure}[t]
\includegraphics{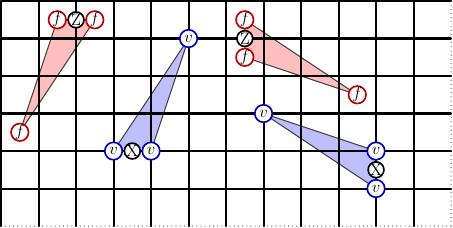}
\caption{\label{Fig:Errors} Single Pauli-$X$ and Pauli-$Z$ errors on horizontal $L$ qubits and vertical $R$ qubits, together with their respective syndromes for the gross code. }
\end{figure}

We can write bit-flip errors on BB codes as a vector of polynomials
$e = \left( e_1 ~~ e_2 \right)^T,$
with polynomial components denoting the Pauli-$X$ error on the left and right qubits respectively. We therefore obtain the syndrome acting on the Pauli-$Z$ stabilizers
\begin{equation}
s = \sigma^T e  = \sum_{j,k} s_{j,k} x^j y^k ,
\end{equation}
where $s_{j,k} = 1$ if and only if the vertex stabilizer at site $(j,k)$ is violated. In Fig.~\ref{Fig:Errors} we show example errors with their respective syndromes for the gross code.

\subsection{Code symmetries}
\label{SubSec:CodeSymmetries}

A symmetry of a code~\cite{Brown2020parallelized, Brown2022conservation} is a subset of its stabilizers that are constrained such that any physical error necessarily violates an even number of stabilizers of the symmetry. In other words, the subset of stabilizer generators that form the symmetry multiply together to give the identity element.
As we will explain in \cref{Sec:Decoder}, this property enables the use of matching on the symmetry. 
The notion of a symmetry has also been generalized to account for biased noise~\cite{Tuckett2020fault, BonillaAtaides2021XZZX, Srivastava2022xyzhexagonal, Miguel2023cellularautomaton, Huang2023tailoring, Sahay_2023Tailoring}.

Once again, we will concentrate only on Pauli-$Z$ type stabilizers, $\sigma$ of \cref{Eqn:Pauli-$Z$stabilizers}.
In the language of polynomials, we define a symmetry $\Sigma$ as follows:
\begin{equation}
  \sigma  \Sigma = 0, \label{Eqn:SymmetryDef}
\end{equation} 
where $\Sigma$ is a polynomial representing the site locations for each stabilizer of the symmetry.

We discover symmetries for BB codes by performing Gaussian elimination on the code check matrix. As we describe in App.~\ref{SubSec:GaussianElimination}, after Gaussian elimination, the last few rows of the transformation matrix for row reduction produce sums of stabilizers that add to zero. These sums exactly form code symmetries. Note that symmetries form an Abelian group by addition modulo 2, i.e., if we have polynomials $\Sigma_j$ and $\Sigma_k$ satisfying $\sigma\Sigma_j = \sigma\Sigma_k = 0$, then we have that $\sigma(\Sigma_j+\Sigma_k) = 0$ by linearity.

\begin{figure}
\includegraphics{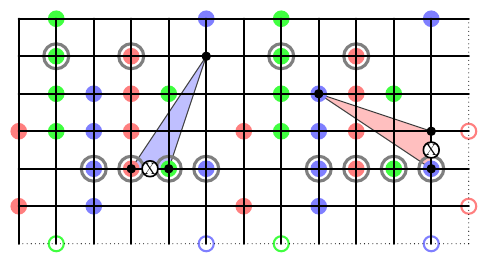}
\caption{\label{Fig:Symmetry} The vertex stabilizers of a BB144 symmetry are marked by red, green and blue dots. All errors violate an even parity of symmetry stabilizers, as shown by $L$ and $R$ errors. The symmetry can be decomposed into right subsymmetries, which are given different colors. The right error violates two vertices of the blue subsymmetry. A left subsymmetry is also shown by dots enclosed by circles. }
\end{figure}

\ex{BB144} We now describe a code symmetry of the gross code under code-capacity bit-flip noise using the polynomial formalism.
\begin{equation}
\Sigma = (x + y^2 +x^2y) (1+x^6) (1 + y^2) A. \label{Eqn:Symmetry}
\end{equation}
where $A$ is the polynomial in \cref{Eqn:AandB}. In Fig.~\ref{Fig:Symmetry}, we show the subset of stabilizers forming $\Sigma$ using colored spots on the vertices of the square lattice. Note that this symmetry consists of $36$ out of the $72$ stabilizer generators. We leave it to the reader to verify that Eqn.~(\ref{Eqn:SymmetryDef}) holds for this choice of $\Sigma$. See App.~\ref{App:SymPics} for more examples.

The gross code has 6 linearly independent symmetries. A generating set of these symmetries can be obtained from distinct translations of $\Sigma$, such that:
\begin{equation}
\begin{aligned}
\Sigma_1 = \Sigma, \quad &\Sigma_2 = y \Sigma, \quad \Sigma_3 = x^3 \Sigma,\\
\Sigma_4 = x^2y^3 \Sigma, \quad &\Sigma_5 = x^3y^3 \Sigma, \quad \Sigma_6 = x^4 y^{-1} \Sigma.
\end{aligned}  
\label{eq:transSym}
\end{equation}

\subsection{Subsymmetries}
\label{SubSec:Subsymmetries}

We define a subsymmetry as a symmetry with respect to a certain subset of errors~\cite{Tuckett2020fault}. By definition, any error generated from this specific subset of errors will always create an even parity of stabilizer violations on the stabilizers of the subsymmetry. 

For BB codes, we choose to look for subsymmetries for errors on only the $L$ qubits, $\Sigma^L$, and subsymmetries for errors on only the $R$ qubits, $\Sigma^R$.
For a $\Sigma^L$ subsymmetry in a BB code,
\begin{equation}
\Sigma^L \sigma = 
\begin{pmatrix} 
0 \\ 
\Sigma^L B
\end{pmatrix}. \label{eq:subsym}
\end{equation}
Observe that the first element of this vector is 0. This means that $\Sigma^L \sigma$ has no support on any $L$ qubits, i.e. it is insensitive to any errors on the $L$ qubits. In practice, this means that any configuration of errors on $L$ qubits will always violate an even number of stabilizers at the locations specified by $\Sigma^L $. 
We can regard subsymmetries as meta-stabilizers $\Sigma^L \sigma$ that detect errors that do not respect the subsymmetry.
The definition of $\Sigma^R$ follows by analogy.

\ex{BB144} Here, a $\Sigma^R$ subsymmetry is:
\begin{equation}
\Sigma^R = (1+x^6) (1 + y^2) B,
\end{equation}
where $B$ is the polynomial in \cref{Eqn:AandB}. 
 We can also express \cref{Eqn:Symmetry} as follows:
\begin{equation}
\Sigma = (y^2 + x +x^2y) \Sigma^R.
\end{equation}
We can thus generate a symmetry using  $ (x + y^2 +x^2y) $ translations of $\Sigma^R$, where each monomial translation forms a distinct linearly-independent subsymmetry. 
Similarly, we can write  $\Sigma^L$ as:
\begin{equation}
\Sigma^L = (1+x^6) (1 + x^2) A.
\end{equation} 

We show the three distinct $\Sigma^R$ subsymmetries that make up $\Sigma$ in Fig.~\ref{Fig:Symmetry} by giving the vertices of different subsymmetries different colors. A single error on an $R$ qubit, shown in red, respects these subsymmetries, i.e. it always violates an even parity of stabilizers for each subsymmetry: no stabilizers of the red or green subsymmetries, and  two stabilizers of the blue subsymmetry. On the other hand, the error on the $L$ qubit violates the red and green subsymmetries by producing one stabilizer violation on each. 
We show an example of $\Sigma^L$, respected by errors on the $L$ qubits, with circled vertices in Fig.~\ref{Fig:Symmetry}.

\subsection{Equivalence between codes on a two-dimensional lattice}
\label{SubSec:LocalEquivalence}

Essential to our decoder is the fact that any stabilizer code in two dimensions with a growing distance is equivalent to some number $K$ copies of the toric code~\cite{Bombin2012universal,Bombin2014structure}. Here, we review this equivalence. There is a finite depth quantum circuit with finite-range gates that maps the given code to copies of the toric code and decoupled ancilla qubits, i.e.,
\begin{equation}
U \left( \mathcal{S}\otimes \mathcal{A} \right) =  \mathcal{S}_\textrm{t.c.}^{\otimes K}\otimes \mathcal{A}'. 
\end{equation}
Here, $U$ denotes a local constant-depth circuit, $\mathcal{S}$ is the stabilizer group of the two-dimensional code of interest, $\mathcal{S}_{\textrm{t.c.}}^{\otimes K}$ is the stabilizer group for $K$ copies of the toric code, and $\mathcal{A}$ and $\mathcal{A}'$ denote an ancilla system in a product state, where some constant number of ancilla qubits are placed at each lattice site of the code.

The fact that $U$ acts locally on the lattice and has constant depth $\Delta$ means that codes $\mathcal{S}$ and $\mathcal{S}_\textrm{t.c.}^{\otimes K}$ share many common properties. For instance, the local toric code stabilizer $\sigma_\textrm{t.c.}\in \mathcal{S}_\textrm{t.c.}^{\otimes K} \otimes \mathcal{A}'$ has local support on the original code $U^\dagger \sigma_\textrm{t.c.} U \in \mathcal{S} $. Specifically, the support is spread by no more than a constant factor $\Delta$ under the mapping due to the light cone of the circuit $U^\dagger$.

In addition to maintaining the local properties of the codes under the mapping, the syndromes of the  code $\mathcal{S}$ are also commensurate. 
Let us assume a large lattice, such that the dimensions $M$ and $N$ of the lattice are much larger than the size of the support of a stabilizer, $r$. In the toric-code picture, errors are string-like objects that give rise to violated stabilizers in pairs. Given some string-like Pauli operator $C$ that corrects a pair of violated stabilizers in the toric code picture, we find an operator $U^\dagger C U$ with ribbon-like support that acts on code $\mathcal{S}$ that will reverse the equivalent syndrome. Again, the width of the ribbon like support is at most $O(\Delta)$.

The fact that errors in the toric code give rise to pairs of violated stabilizers at the end points of a string-like error follows immediately from the physics of the toric code. Specifically, the syndrome must respect symmetries of the code, and for the toric code, a symmetry is composed of all the stabilizers~\cite{Brown2020parallelized,Brown2022conservation}. The symmetries of the toric code $\mathcal{S}_\textrm{t.c.}^{\otimes K} \otimes \mathcal{A}'$ must also carry over to $\mathcal{S}$. This is verified using the unitarity of $U$. Indeed, for a symmetry $\Sigma' \subset \mathcal{S}_\textrm{t.c.}^{\otimes K} \otimes \mathcal{A}$ such that $\prod_{\Sigma'} \sigma_\textrm{t.c.} = 1$ we have
\begin{equation}
\prod_{\Sigma'} U^\dagger \sigma_\textrm{t.c.} U = U^\dagger\left(\prod_{\Sigma'}  \sigma_\textrm{t.c.}\right) U = U^\dagger U =1, 
\end{equation}
when all elements of $\sigma_\textrm{t.c.} \in \Sigma'$ are conjugated by unitary operator $U^\dagger$ to transform the terms onto stabilizers of $\mathcal{S}$. We can therefore adapt conventional strategies for decoding the toric code based on code symmetries to bivariate bicycle codes given a sufficiently large lattice.

 For a family of two-dimensional translationally invariant (2DTI) quantum LDPC codes, described by a fixed polynomial matrix, there is a constructive algorithm to generate a local unitary mapping to copies of the toric code in its standard presentation~\cite{haah2016algebraic,Haah2021classification,PRXQuantum.5.030328}. 
This local unitary equivalence requires blocking many sites together and so breaks the translation invariance of the original code down to a subgroup. 
However, the equivalence to toric codes is already present in the original model without explicitly applying a local unitary or breaking the translation symmetry.  Preserving the translation symmetry results in an equivalence to translation-symmetry-enriched toric codes, where the $\mathbb{Z}\times \mathbb{Z}$ translation group may act nontrivially by permuting the anyon types; see Appendix~\ref{app:TDN}. This is referred to as weak symmetry-breaking in Ref.~\cite{kitaev2006anyons}. 
A procedure to calculate the action of these translation operators on the toric code anyon theories is described in Appendix~\ref{app:TCE}, and Ref.~\cite{Dua2019}.

\begin{figure}[t]
    \centering
    \includegraphics[page=3]{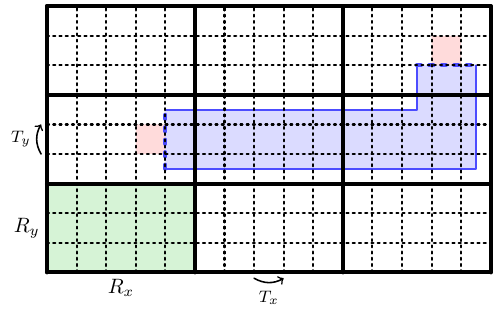}
    \caption{The unfrustrated $R_x \times R_y$ unit cell (green) superimposed onto the original lattice (dotted) with $T_x$ and $T_y$ translation generators. 
    A ribbon operator (blue) creates a pair of syndromes (red) that are related by an unfrustrated translation $T_x^{2 R_x}T_y^{R_y}$. }
    \label{fig:UnfrustratedLattice}
\end{figure}

We denote the generator of translation along the $x$-axis by $T_x$, and similarly translation along the $y$-axis by $T_y$. 
The order of the $T_x$ translation action on the anyons is denoted $R_x$, and similarly the order of $T_y$ is denoted $R_y$. 
These orders must be finite, as the number of anyons and the size of their automorphism group is finite. 
However, they may be exponentially large in the number of anyon types~\cite{Dua2019}.
The orders set the size of an $R_x \times R_y$ \textit{unfrustrated} unit cell, see Fig.~\ref{fig:UnfrustratedLattice}. 
Breaking the translation group down to the subgroup generated by $T_x^{R_x},$ $T_y^{R_y}$, which translates unfrustrated unit cells into one another, results in a trivial action on the anyon types. 
This sets the scale for the coarse-grained local unitary equivalence of the 2DTI to copies of toric code with standard checks. 
It also sets the scale for the existence of ribbon operators that create a syndrome, together with a translation of that syndrome by some $T_x^{n R_x} T_y^{m R_y}$, for $n,m\in\mathbb{N}$. 

Finally, an unfrustrated $R_x \times R_y$ torus must contain $2^{2 K}$ symmetries corresponding to parity conservation of each anyon type in the equivalent copies of toric code.
These symmetries can be tiled on the infinite plane to produce all nontrivial global symmetries of the original infinite 2DTI code. 
Hence, all the symmetries of a 2DTI code can be viewed as descending from the conventional symmetries of toric codes with frustrated boundary conditions, see Appendix~\ref{app:Unfrustrated}.
Here, we use these symmetries to decode 2DTI codes, with a special focus on BB codes.

\section{Decoder}
\label{Sec:Decoder}

In this section, we lay out our decoding scheme. In \cref{SubSec:SyMatchGraph}, we start  with a review of the matching algorithm before describing how one can generate an input graph for matching from a code symmetry. After matching on a symmetry, a correction can be found from the results of matching using the `cylinder trick', which we propose in \cref{SubSec:Cylinder}. 

Matching on symmetries has previously been demonstrated with topological codes~\cite{Dennis02, Sahay2022decoder, Benhemou2025minimising}, fracton codes~\cite{Brown2020parallelized, SchwartzmanNowik2025generalizing} and classical fractal codes~\cite{Nixon2021correcting}. Our strategy of matching on symmetries for BB codes, hereon referred to as the \textit{symatch} decoder, functions well for a subclass of BB codes. However, for certain BB codes, symatch can fail for low-weight errors. To address this, we introduce additional subroutines in \cref{SubSec:Simplex} and \cref{SubSec:CorrSubSym} to improve its logical error rate. These respectively involve using consistency conditions from matching results on an overcomplete set of code symmetries, and the use of subsymmetries to identify and correct $L$- or $R$-only error configurations with fast classical decoders.

\subsection{Constructing a symmetry graph for matching}
\label{SubSec:SyMatchGraph}

The minimum-weight perfect matching (MWPM) algorithm~\cite{Edmonds1965MaximumMA, Higgott2023improved,wu2023fusionblossomfastmwpm} takes as input a weighted graph $\mathcal{G} = (\mathcal{V}, \mathcal{E})$, in which there are an even number of highlighted vertices $\mathcal{D} \subseteq \mathcal{V}$. Note that any edge $e$ in $\mathcal{G}$ must connect exactly two vertices, that is $|e| = 2, \forall e$. The  algorithm returns a pairing of elements of $\mathcal{D}$ along a subset of edges $\mathcal{E}' \subseteq \mathcal{E}$ such that the total path length between pairs along $\mathcal{E}'$ is minimized. 

We can construct $\mathcal{G}$ for QECCs for the purposes of decoding. A correction can be extracted from the edge subset $\mathcal{E}'$ returned from matching on $\mathcal{G}$. 
Note that though we focus on MWPM in this work, this task of `matching' can be implemented using any method considered for the toric code.
This includes clustering and union-find~\cite{Bravyi2013quantum,Anwar2014fast,Delfosse_2021, Huang_2020} which, similar to matching, pair up local violated stabilizers.

\begin{figure}[t]
    \centering
    \includegraphics[width=0.9\linewidth]{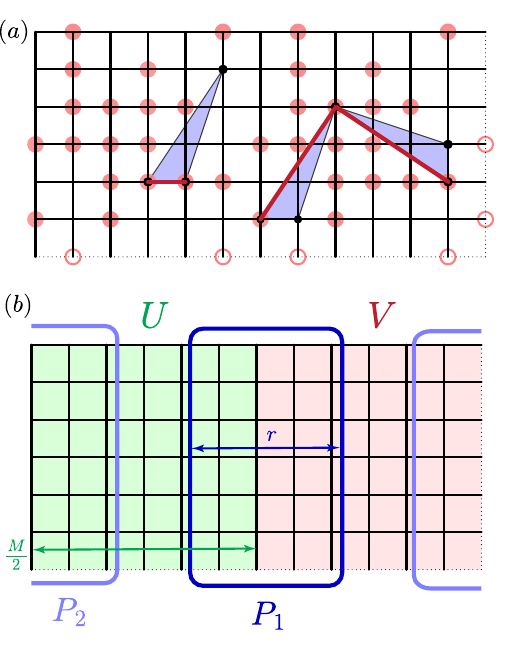}
    \caption{(a) Constructing a matching graph from a symmetry. We show three errors. Edges are drawn between stabilizers in the symmetry that are violated by these errors. (b) We geometrically divide symmetries into cylinders into $U$ and $V$ with width $M/2 \gg$ $r$, the stabilizer span. The product of Pauli operators of the symmetry supported within $U$, isolated along one cylinder boundary, forms a non-trivial logical loop. }
    \label{fig:DecTricks}
\end{figure}

We now explain how to generate the weighted graph with $|e| = 2, \forall e$ required by MWPM from a symmetry $\Sigma$ of a BB code. We term this the symmetry graph, $\mathcal{G}_\Sigma$. 
This graph has a vertex set $\mathcal{V}$ such that each vertex corresponds to a stabilizer generator of the symmetry. We define edges of the symmetry graph, under the assumption of an independent and identically distributed bit-flip noise model, as follows: We consider introducing an individual bit-flip at each qubit of the lattice, one by one. We then draw an edge between each pair of stabilizers of the symmetry that the error has violated, see \cref{fig:DecTricks}(a). 
A symmetry guarantees that all individual errors must create an even number of defects. 
In the case that an error violates a large, necessarily even, number of stabilizers of the symmetry, we can draw a complete set of edges between all pairs. Repeating this for each individual error gives us the complete symmetry graph.

For i.i.d. bit-flip noise, all edges are assigned the same weight. We can generalize the weight function for the edges of $\mathcal{G}_\Sigma$ to aid decoding. 
Here, some of our decoder variants use belief propagation (BP)~\cite{Kschischang_2001,Poulin_2008} to reweigh edges of the symmetry graph via message passing. Note that we perform BP using the complete syndrome, including violated stabilizer generators that are not elements of the symmetry of interest. 

Given a symmetry $\Sigma$ and its  derived weighted symmetry graph $\mathcal{G}_\Sigma$, we can perform matching on $\mathcal{G}_\Sigma$ for a given syndrome, i.e. highlighted vertex subset $\mathcal{D} \subset \mathcal{G}_\Sigma$. The matching algorithm will return a set of edges pairing up these vertices with a minimum total weight.

\subsection{The cylinder trick}
\label{SubSec:Cylinder}

Here, we describe the `cylinder trick' essential to our decoder. Performing the trick on a symmetry $\Sigma$ extracts a logical operator $\overline{L}$ associated with $\Sigma$. 
As we explain below, this enables us to interpret the results of matching on $\Sigma$ to find a correction for ${\overline{L}}$. 
The cylinder trick has been described explicitly for the color code in Appendix~A of Ref.~\cite{Sahay2022decoder}, but we find the trick works generically for 2DTI codes such as the BB codes considered here. We  break down our description of this method as follows:

\subsubsection{Extracting a logical operator from a symmetry}

We start with $\Sigma$ on a toric manifold. We first take the stabilizer generators included in $\Sigma$ and divide them into two disjoint subsets $\Sigma^U$ and $\Sigma^V$ according to their location such that $\Sigma = \Sigma^U \cup \Sigma^V$. For simplicity, we take $\Sigma^U$ and $\Sigma^V$ to be supported on two cylinders $U$ and $V$, and we assume that both cylinders are large relative to the size of the support of a stabilizer. As an example, for a lattice of dimensions $ M \times N$ with stabilizers assigned to sites $(x,y)$ with $0\le x < M$ and $0 \le y < N$, we might say $\Sigma^U$ includes all stabilizers of $\Sigma$ on sites of $U$ with $0 \le x < M/2 $ and $\Sigma^V$ includes all other stabilizer generators of $\Sigma$ with $M/2 \le x < M$, see \cref{fig:DecTricks}(b). We take $M/2 \gg r$ where $r$ is the diameter of a box that contains the support of a stabilizer operator on our lattice.

Let us look more closely at these two disjoint subsets of $\Sigma$. We can write down an operator
\begin{equation}
P = \prod_{\sigma\in\Sigma^U} \sigma = \prod_{\sigma\in\Sigma^V} \sigma, \label{Eqn:Cylinder}
\end{equation}
where $P$ is an element of the stabilizer group supported near the two boundaries of the cylinders supporting the $\Sigma^U$ generators and $\Sigma^V$ generators. 
We can express $P = P_1P_2$ as the product of two disjoint operators $P_1$ and $P_2$, where $P_1$ and $P_2$ are supported close to the opposite boundaries of the cylinders. If $M/2 \gg r$, these disjoint components are well separated. Consequently, any stabilizer generator of diameter $r$ can share support with at most one of either $P_1$ or $P_2$.  

$P = P_1P_2$ is a member of the stabilizer group that commutes with all stabilizer generators. It follows that, due to their disjoint support, $P_1$ and $P_2$ must individually commute with all elements of the stabilizer group. Operators that commute with all elements of the stabilizer group are either themselves stabilizers or are logical operators. For a symmetry corresponding to a nontrivial anyon type, $P_1$ and $P_2$ must be nontrivial logical operators, see App.~\ref{app:Unfrustrated}. 
Intuitively, each has the support of a string-like operator that wraps around a non-trivial cycle of the lattice embedded on a torus. We choose ${\overline{L}}=P_1$.

\subsubsection{Determining a correction}
 \label{SubSec:Symbl}

At this stage, we briefly digress to describe a general prescription for decoders to determine corrections. In its simplest terms, a decoder needs only to estimate the parity of errors on a generating set of logical operators. Given an error $E$, we must estimate the commutators $b_{\overline{L}} =[E,{\overline{L}}]$ for each generating logical operator ${\overline{L}}$. Having obtained $\{b_{\overline{L}}\}$, we may simply propose some initial correction $C'$ which is readily achieved by using the destabilizers~\cite{aaronson_2004improved} of the code. We then take this initial correction and find the true correction $C =  {\overline{L}}'C'$ where ${\overline{L}}' $ is some logical operator we choose such that $[C,{\overline{L}}] = [E,{\overline{L}}]$ for all ${\overline{L}}$. 

Having extracted ${\overline{L}} = P_1$ from $\Sigma$ above, we  now describe how we determine $b_{\overline{L}}$  using matching on $\mathcal{G}_{\Sigma}$. 
We start off by considering a simplified noise model where errors only happen on the support of ${\overline{L}}$. 
Using the definitions given above, we argue that an error $E$ that commutes (anticommutes) with ${\overline{L}}$ must necessarily produce a syndrome with an even (odd) parity of stabilizer generators in both subsets $S\in\Sigma^U$ and $S\in\Sigma^V$. It follows that
\begin{equation}
b_{\overline{L}} = (-1)^{|\mathcal{E}'|}, \label{Eqn:EdgesInterpretation}
\end{equation}
where $|\mathcal{E}'|$ denotes the number of edges returning from matching on $\mathcal{G}_{\Sigma}$ with one terminal vertex on $U$ and one terminal vertex on $V$, such that the edge crosses ${\overline{L}}$.

Let us prove ~\cref{Eqn:EdgesInterpretation}. For the case that $E$ commutes (anticommutes) with ${\overline{L}}$, the stabilizer element $P$ as defined in \cref{Eqn:Cylinder} commutes (anticommutes) with $E$. Given that $P = +1$ ($P=-1$) is the product of local generators on the cylinder $\Sigma^U$ and $\Sigma^V$, then it follows that there must be an even (odd) parity of generators $S\in\Sigma^U$ and $S\in\Sigma^V$ in order for \cref{Eqn:Cylinder} to hold. Given that there is necessarily an even (odd) number of vertices on each cylinder for the case where $E$ commutes (anti commutes) with ${\overline{L}}$, then a complete matching over $\Sigma$ must produce an even (odd) number of edges crossing from cylinder $U$ to cylinder $V$, thereby verifying~\cref{Eqn:EdgesInterpretation}.

Let us remark that at no point in our argument have we assumed that an error must give rise to only two violated stabilizers for a given symmetry. 
All that is assumed is that an error will violate an even number of symmetry stabilizers, which is true due to the structure of the symmetry itself. 
Decoding using the cylinder trick extends to such cases where we decompose higher-weight hyperedges of the matching hypergraph into, say, a complete set of edges between all pairs of vertices for a given hyperedge. Beyond this change, all of our arguments presented above will follow in producing a matching decoder.

\subsubsection{The cylinder trick for small codes}

Thus far, it appears that our extraction of ${\overline{L}}$ depends on the ability to find cylinders $U$ and $V$ on the code manifold such that the width of these cylinders is much larger than the geometric support of a stabilizer generator. Unfortunately, this property is not satisfied for several practical codes with a small number of qubits $n$. Here, we show that, even for these small codes, we can effectively use the cylinder trick. 

We start with a BB code $\mathcal{S_{\textrm{base}}}$ with lattice dimensions $(M,N,\alpha)$. Let us assume, without loss of generality, that  $M$ is the limiting dimension $M /2 \leq r$. In this case, we can construct a new code $\mathcal{S_{\textrm{doubled}}}$  with  the same polynomials, but lattice dimensions $(2M,N,\alpha)$. 
We can extract symmetries and their corresponding logical operators on $\mathcal{S_{\textrm{doubled}}}$ using the cylinder trick, or recursively move to an even larger lattice if required. 
When folded back down onto  $\mathcal{S_{\textrm{base}}}$ via modulo two addition, such that
\begin{equation}
q_\textrm{base} = \sum_{j,k \in \textrm{doubled}} q_{j,k} x^{j \mod M} y^k,\label{eq:fold}
\end{equation} 
these logical operators will either remain logical degrees of freedom, or multiply into stabilizers, dependent on the degree of topological frustration of the lattice. This allows the extraction of ${\overline{L}}$ and $\Sigma$ on $\mathcal{S_{\textrm{base}}}$.
We can in fact decode errors on $\mathcal{S_{\textrm{base}}}$ directly on the symmetries of  $\mathcal{S_{\textrm{doubled}}}$: For any error producing a syndrome $s$ on $\mathcal{S_{\textrm{base}}}$, we duplicate $s$ onto $\mathcal{S_{\textrm{doubled}}}$ using 
\begin{equation}
s_\textrm{doubled}  = \sum_{j,k \in \textrm{base} } s_{j,k} \left(x^j y^k +  x^{j+M} y^k \right).
\end{equation}
We then decode on $\mathcal{S_{\textrm{doubled}}}$, and extract $b_{\overline{L}}$ for logical operators of $\mathcal{S_{\textrm{base}}}$ found using \cref{eq:fold}.

\subsubsection{The basic symatch decoder} \label{SubSec:SyMatching}

We can now put the pieces of our construction together to form a complete decoder based on matching on symmetries: Let a BB code $\mathcal{S}$ have $K$ linearly independent symmetries $\{\Sigma_{1} \cdots \Sigma_K\}$. 
Then, $\mathcal{S}$ has $k = 2K$ independent logical operators. Pairs $({\overline{L}}_{i,h}, {\overline{L}}_{i,v}), i\in \{1, ... K\}$ of logical operators can be found using the cylinder trick horizontally and vertically respectively on $\Sigma_{i}$. It is sufficient to find a correction for this generating set of logical operators by determining the value of $b_{{\overline{L}}_{i, q}}, q\in \{h,v\}$ from matching on the symmetry graph $\mathcal{G}_{\Sigma_i} \forall i$ as outlined in \cref{SubSec:Symbl}. We term this the symatch decoder, shown using dashed blue lines in \cref{fig:codecappl}.

\subsection{Over-matching: the simplex-symatch decoder}
\label{SubSec:Simplex}

 While symatch functions well for a subclass of BB codes (\cref{fig:codecappl} left column), we find that, in the general case, individual matching results on symmetries may underperform (\cref{fig:codecappl} right column) due to short loops in $\mathcal{G}_{\Sigma}$. Here, we attempt to improve the reliability of our decoder using redundant matching subroutines. We generate $2^K-1$ non-trivial symmetries via linear combinations of the $K$ independent symmetries of a code. Given that we can now match on up to $2^K-1$ symmetries, in this section we consider how we might compare the results of this over complete set of matching results to identify incorrect outcomes.

Let us first outline how we can construct constraints among matching results. We consider three symmetries: $\Sigma_1$, $\Sigma_2$ and $\Sigma_{1+2} = \Sigma_1 + \Sigma_2$, where the third constrains the other. By linearity, using the cylinder trick on these three symmetries enables us to find the number of errors on logical operators $\overline{Z}_1$, $\overline{Z}_2$ and $\overline{Z}_1\overline{Z}_2$ respectively, where we choose to cut the cylinder from the same locations on the torus for each symmetry. Each matching result gives a corresponding bit outcome $b_1$, $b_2$ and $b_{1+2}$. Here, assuming no matching results give incorrect results, these bit values are constrained to agree. We have that
\begin{equation}
b_1 \oplus b_2 = b_{1+2},
\end{equation}
where addition is performed modulo 2. Equivalently, we have the check $b_1 \oplus b_2 \oplus b_{1+2} = 0$.
If this check constraint is violated, we can detect a single matching failure. 
In general, we find several constraints among $2^K-1$ matching results. In App.~\ref{App:Hamming}, we explicitly show how we can encode the complete set of $2^K-1$ matching results into a classical simplex code~\cite{eczoo_simplex} with parameters  $[2^K-1, K, 2^{K-1} ]$ to correct upto $K/2$ matching failures, naming this the simplex-symatch decoder, shown via solid blue curves in \cref{fig:codecappl}.

Our proposal of using an overcomplete set of matching subroutines is distinct from previous ideas.
In prior work, matching results have been passed between $X$ and $Z$ matching subroutines on the toric code to exploit $Y$ error correlations~\cite{fowler2013optimalcomplexitycorrectioncorrelated, liu2025correlatedmatchingdecoder488}. 
Additionally, matching problems have been combined on a unified lattice to improve logical error rates~\cite{Sahay2022decoder,Benhemou2025minimising, gidney2023newcircuitsopensource}.
In contrast to Refs.~~\cite{fowler2013optimalcomplexitycorrectioncorrelated, liu2025correlatedmatchingdecoder488}, correlations between bits $b_j$ can be detrimental to our over-matching strategy.  
To see this, consider a contrived example with two  toric codes, with individual symmetries $\Sigma_1$ and $\Sigma_2$, consisting of all the stabilizers of each individual code. 
We can construct the symmetry $\Sigma_{1+2} = \Sigma_1 + \Sigma_2$. Matching on this additional symmetry essentially performs matching on both copies of the toric code in parallel on two disjoint graphs in a single subroutine. Any error that causes a logical failure on $\Sigma_1$, will also cause matching on symmetry $\Sigma_{1+2}$ to fail with certainty. 
As such, with the exception of cases where a matching subroutine is guessing between two logically inequivalent weight $d/2$ errors, over-matching offers little advantage.

In general, we do not find highly correlated failures between matching subroutines. 
We make two further hypotheses as to why simplex-symatch offers improvements: First, we note that matching on any given symmetry will neglect some syndrome information. Redundant over-matching allows all of the syndrome data of the code to be considered more uniformly. Secondly, it is likely that a large fraction of individual matching subroutine failures are cases where our consistency conditions among redundant matching subroutines probabilistically detect a failed matching result. As we increase the number of matching subroutines, the likelihood of a consistency check flagging these errors will increase. 

In future work it will be interesting to establish the conditions under which over-matching offers an advantage over using a minimal symmetry set. 
It is also worth commenting on the fact that this decoding step does not scale favorably in the number of logical qubits $k = 2K$, at least, not while using the simplex classical code with $2^{K}$ matching subroutines.  
It may be interesting to investigate if it is sufficient to use $<2^K$ redundant matchings  in a smaller outer classical matcher-correcting code.  
For the codes we are considering with relatively small $n$ and $k$, we find that simplex-symatch is quite tractable.

\subsection{Identifying L- and R-only errors using subsymmetries}\label{SubSec:CorrSubSym}

We now describe how we can use subsymmetries to identify and correct $L$-only or $R$-only errors.
Considering only Pauli-$Z$ stabilizers, a BB code maps to copies of classical code $A$ on the left qubits and copies of classical code $B$ on the right qubits. 
For general errors, the stabilizers of the code combine these two classical codes in such a way that we cannot regard the quantum code as two separate classical codes. 
However, for error configurations that are supported either exclusively on the left qubits, or exclusively on the right qubits, we can adopt a fast classical decoder for code $A$ or $B$ respectively. 

We now describe this process for $L$-only errors. These errors must respect $\Sigma^L$, the symmetries of code $A$, see \cref{eq:subsym}. By definition, $\Sigma^L$ is a code subsymmetry. 
Note that errors on $R$ qubits may violate these subsymmetries, creating an odd parity of syndrome defects. Importantly, if our error syndrome respects all $\Sigma^L $ subsymmetries, then we can guarantee that there is a valid correction that is supported entirely on $L$ qubits. 

If all $\Sigma^L $ are valid, we can apply a fast classical decoder on the code $H^Z = [A|0]$. We accept the decoder's correction if its weight is less than $d/2$. If the $L$-only correction is higher-weight, we fall back to matching to identify a lower-weight $LR$ error. In this work, we use BP as our fast classical decoder, and study appending this subsymmetry check and `$L/R$' BP to simplex-symatch as a pre-processing subroutine.

\section{ Results} \label{Sec:Num}

We study bivariate bicycle codes using our symatch decoder variants described in \cref{Sec:Decoder}. The specific code instances we examine are listed in~\cref{table:codes}, and consist of members of the Gross code family,  generalized toric codes, cyclic hypergraph-product codes, and directional codes. We compare the logical error rates (LERs) of our decoder to BP-OSD0~\cite{Roffe_2020} and the tesseract decoder~\cite{beni2025tesseractsearchbaseddecoderquantum,grbic2026acceleratingtesseractdecoderquantum}. We additionally exhaustively study low weight errors for the gross code in \cref{table:grossexh}.

Fig.~\ref{fig:codecappl} contains results of logical error rate versus physical error rates for these codes under code-capacity bit-flip noise. We remark that the baseline comparative of BP-OSD0 (shown in yellow) is not guaranteed to be  \textit{distance preserving}, where we use this term to denote that the decoder only fails for weight $w \geq d/2$ errors. A distance-preserving decoder will show LER scaling $P_L \propto p^{d/2}$. Note that the displayed `long-beam' variant of tesseract (green) has previously been shown to achieve lower LERs than BP-OSD with higher runtime~\footnote{In Fig.~\ref{fig:codecappl}, we use the \texttt{tesseract-long-beam} variant for codes on the right, and \texttt{tesseract} for codes on the left.}.

First, let us discuss the blue curves of \cref{fig:codecappl}, which show the performance of bare symatch and simplex-symatch. For directional codes, the color code, and the cyclic La-cross code shown in \cref{fig:codecappl}(left), symatch achieves $P_L \propto p^{d/2}$. Indeed, for the color code, symatch reduces to the projection decoder~\cite{Delfosse_2014decoding, Kubica_2015unfolding, Kubica_2023efficient}. For the non-topological codes, we highlight that, to our knowledge, this is the first demonstration of a matching decoder. 

\begin{figure}[t]
    \centering
\includegraphics[width=0.49\linewidth]{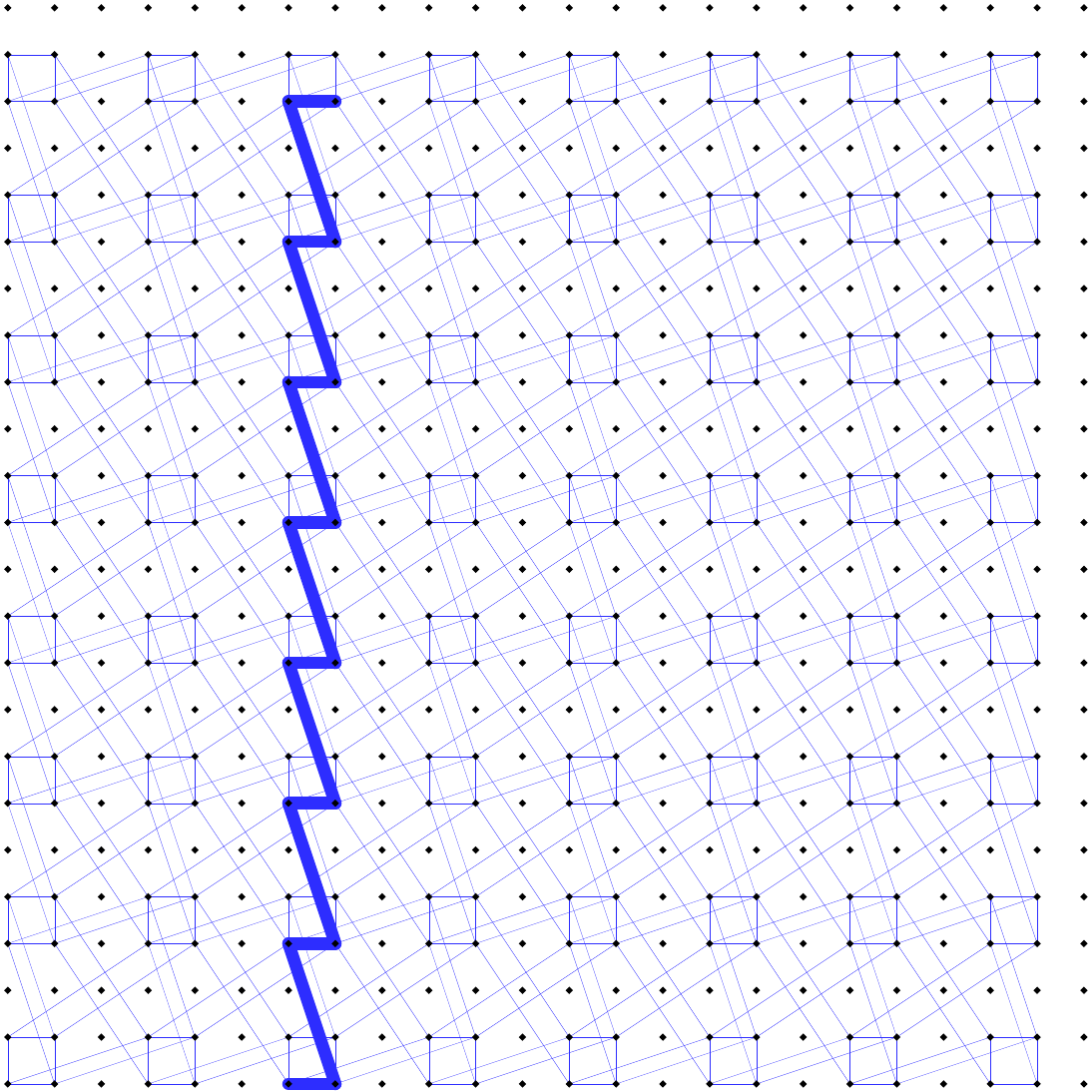}
\includegraphics[width=0.49\linewidth]{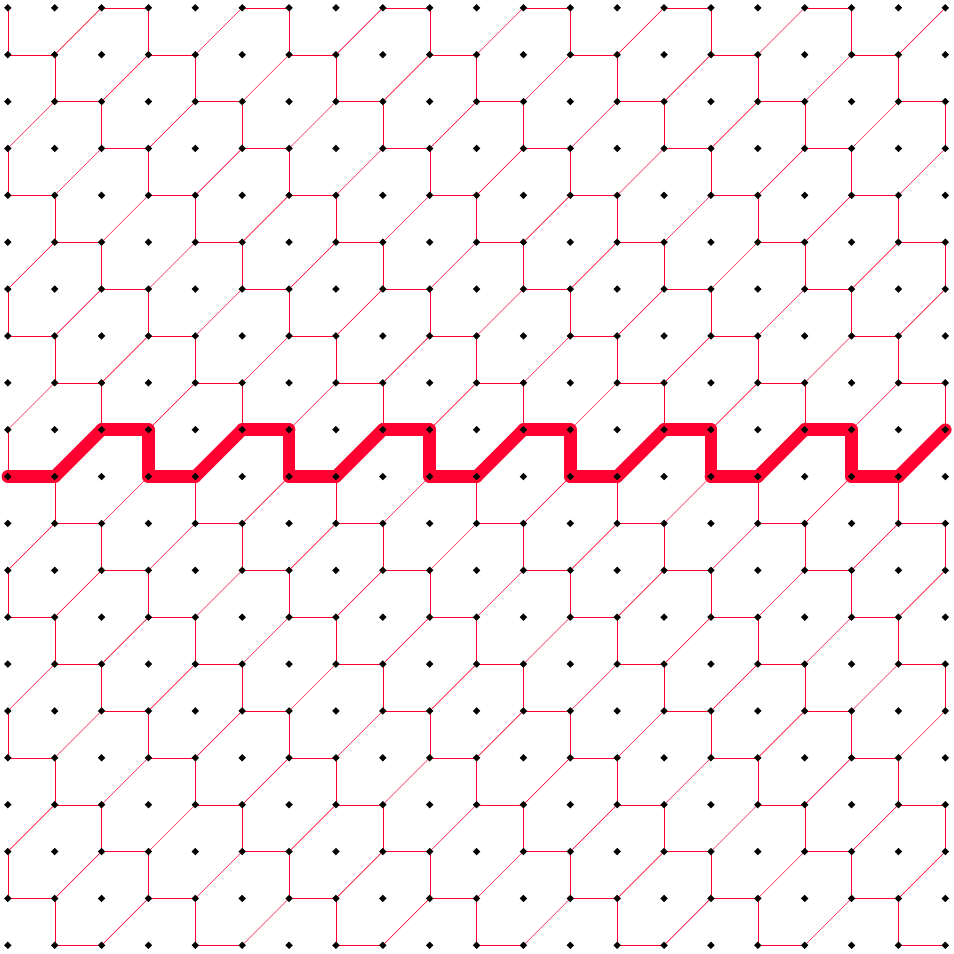}
\includegraphics[width=0.99\linewidth]{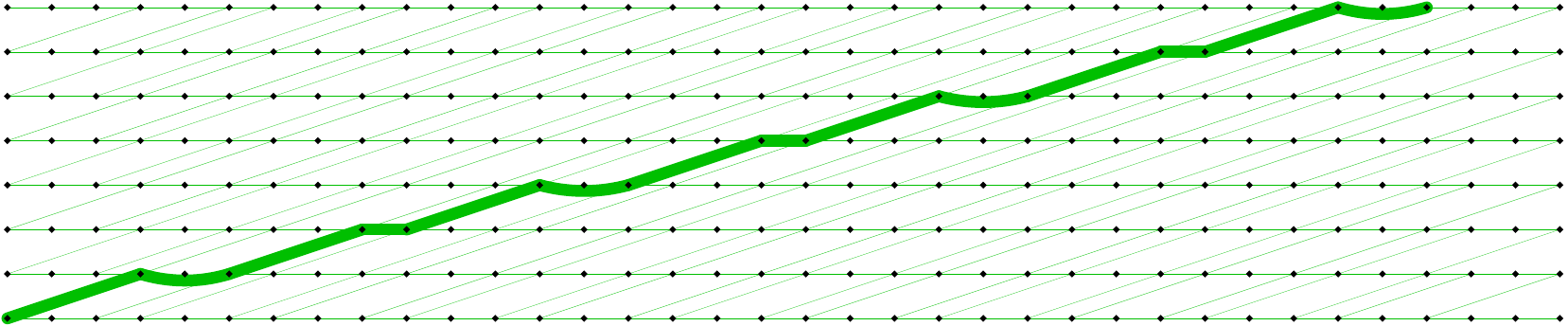}
    \caption{The structure of symmetry matching graphs for the gross code family (left), the toric color code (right), and a $NE^3N$ directional code (bottom). The length of a minimum-weight nontrivial loop, highlighted across the lattice in each subfigure, predicts the effective distance of bare symatch.
    }
    \label{fig:bb16_mg}
\end{figure}

\begin{figure}[t]
    \centering
    \includegraphics[width=0.92\linewidth]{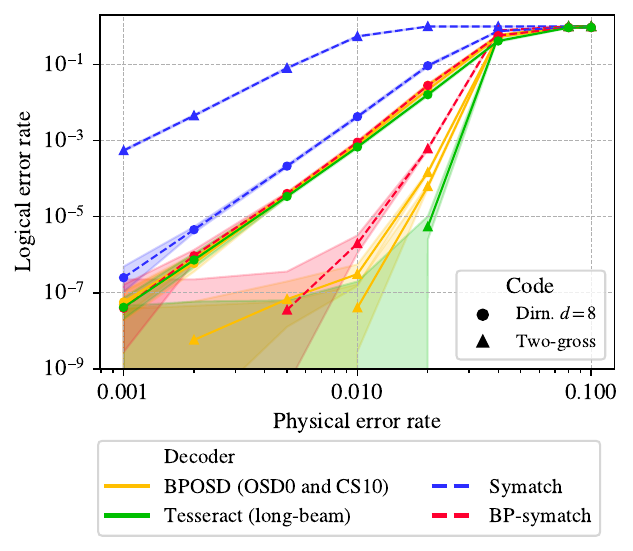}
    \caption{Decoder performance under phenomenological noise for the two-gross code and a $NE^3N$ directional code.}
    \label{fig:phenom}
\end{figure}

For the BB codes in Fig.~\ref{fig:codecappl}(right), symatch is not distance-preserving: The LER scaling of the corresponding blue curves  is limited by weight $w < d/2$ errors. We suggest that this is due to the structure of the underlying torii. 
Specifically, for a code with distance $d$, each $\Sigma$ maps to a toric graph $\mathcal{G}_\Sigma$ with the shortest non-trivial loop of length roughly $d/r$, where $r$ is the size of the support of a stabilizer generator on the lattice. 
In \cref{fig:bb16_mg}, we present $\mathcal{G}_\Sigma$ for $M\times N$-sized codes with large $n$ and omit matching edges on the boundaries. 
\cref{fig:bb16_mg}(left) shows the symmetry matching graph structure for an instance of the gross code family. 
Observe the periodic structure resulting in an effective 6-regular torus of size $2M/3$. This is in line with our symatch results reported in blue in \cref{fig:codecappl} and \cref{fig:phenom}: The LER scales as $p^{d_\Sigma/2}$ where  $d_\Sigma = 8 = 2/3 \times (M=12) $ for the two-gross code. On the right, we show $\Sigma$ for the color code, verifying that $d_\Sigma = 4M/3 $, and below that display $\Sigma$ for a $NE^3N$ directional code.

For these codes with ${d_\Sigma} < d$, simplex-symatch has lower LERs and slightly improved scaling over basic symatch, but low-weight errors may still probabilistically fail all the simplex code consistency checks. The performance of symatch improves dramatically as soon as we introduce BP as a pre-processing step to reweigh edges in the symmetry graphs, shown by the red curves in ~\cref{fig:codecappl}. With the addition of several BP rounds to reweigh the edges of $\mathcal{G}_\Sigma$ based on the complete syndrome for matching, symatch is now competitive with other LDPC decoders for these BB codes. 
Interestingly, the improvement offered by simplex post-processing is largely washed out by the addition of BP at these physical error rates.

We briefly comment on the utility of BP: As we have alluded to earlier, for general BB codes, individual matching subroutines can fail in the presence of low-weight errors, as these subroutines do not use all syndrome information. One can view simplex post-processing as an external patch for this issue, but without information-sharing between subroutines, the ``correct'' result is still selected out of individually unreliable options.
In contrast to simplex-post-processing, BP allows complete syndrome information to be accessed by each subroutine, increasing their individual reliability. See App.~\ref{App:CorrSym} for further discussion on soft information.

\subsection{Phenomenological noise}

We further test our decoder using a more realistic noise model -- the inclusion of  unreliable stabilizer measurements~\cite{Brown2020parallelized, Brown2022conservation}. 
In this case, we replace stabilizers with corresponding detectors over time~\cite{Derks_2025designing}. We recover symmetries in the spacetime model by choosing a subset of detectors that correspond to the stabilizers of a symmetry over all timesteps. This ensures that both bit-flip and measurement errors violate an even parity of detectors.
For a given symmetry in this spacetime model, built from detectors, we generalize the cylinder trick by evaluating changes in the logical operator that corresponds to the symmetry over time. As proof of principle, we present phenomenological data for bare symatch and belief-symatch in \cref{fig:phenom}, with trends mirroring results for code-capacity noise.

\begin{table*}[t]
\centering
\begin{tabular}{|l|rlrlrlrl|rlrlrlrl|}
\hline
 & \multicolumn{8}{c|}{Failures on vertical $\overline{L}$ (weight fraction)} & \multicolumn{8}{c|}{Failures on horizontal $\overline{L}$ (weight fraction)} \\
\hline
Decoder  & wt2  & & wt3  & & wt4 &  & wt5 &  & wt2 &  & wt3 &  & wt4 &  &wt5 &\\
\hline
Symatch              &  81 & ($8$e-3) & 19691 & ($4$e-2) &  & &  & & 296 & (3e-2)   & 51771 & ($1$e-1) & &  &  & \\
Simplex-Symatch      &  10  & ($1$e-3) & 2062 & ($4$e-3)  &  &  & & & 0 & & 1121 & ($2$e-3) &  & && \\
L/R Symatch            & 0 & & 13029 & ($3$e-2) & & & &  & 126 & (1e-2)   & 38438 & ($8$e-2) &  & && \\
L/R Simplex-Symatch   &  0 & & 1061 & ($2$e-3) & & &  &  & 0 &  & 822 & ($2$e-3) &  &&&  \\
\hline
BP Symatch           & 0 & & 0 & & 0 &  & 622 & (1e-6) & 0 & & 0 & &  0 & & 1129& (2e-6) \\
BP  Simplex-Symatch  & 0 & & 0 && 0 && 472& (1e-6) & 0 && 0 && 0 && 1078 &(2e-6) \\
BP L/R Symatch      & 0 & & 0 & &  0 & &151 & (3e-7) & 0 & & 0 & &  0 & & 512 & (1e-6)\\
BP L/R Simplex-Symatch & 0 && 0 && 0 && 79& (2e-7) & 0 && 0 && 0 && 462& (1e-6)\\
\hline
BP OSD0            & 0& & 0 & & 0 & & 1071 &($2$e-6) & 0& & 0 && 0 &  & 1068 & ($2$e-6) \\
BP OSD-CS10            & 0 && 0  &&  0&  & 450 &($9$e-7) & 0 && 0 && 0 &  & 501 & ($1$e-6) \\
\hline
\end{tabular}
\caption{We exhaustively count decoder failures for low-weight errors for the gross code. Central columns count failures to correct logical operators found using the vertical cylinder trick given errors of weight wt $\in \{2,3,4,5\}$. Right columns evaluate failures to correct logical operators found using the horizontal cylinder trick. The weight fraction in brackets represents the fraction of total operators of that weight ($= ^{144}C_{\text{wt}}$) that cause a decoder failure, reported upto one significant figure. All BP subroutines used have the hyperparameters \texttt{bpmethod=minsum,maxiters=1000,prior=3/144}.  As we focus on the lowest-weight failing errors for any decoder, we do not report values in the empty cells. }
 \label{table:grossexh}
\end{table*}

\subsection{Exhaustive search for low-weight failures}

We exhaustively enumerate low-weight code capacity errors that lead to decoder failure for different decoders on the gross code in \cref{table:grossexh}.
Here, in addition to the decoders discussed above,
we also use the subsymmetry structure to attempt to isolate and decode $L$-only and $R$-only errors using classical BP subroutines. We observe that BP-simplex-symatch has commensurate performance to BPOSD0  for these low weight errors. Interestingly, the $L/R$ preprocessor offers noticeable improvements in this low-error rate regime: With this protocol, the number of errors causing failures drops by a factor of 20\%-80\%. Crucially,  $L/R$ preprocessing pushes our decoder to have better performance than BPOSD-CS10 for these low-weight errors. Despite these improvements, no decoder listed can correct all weight-$5=\lfloor d/2\rfloor$ errors.

The table also shows a curious asymmetry in decoding the vertical logicals of the gross code over the horizontal logicals using symatch decoders, with the vertical logical operators consistently demonstrating lower failure rates. We attribute this to the small $N$ dimension of the gross code, which means it is much easier for errors to wrap around the torus vertically than horizontally.

\subsection{A comment on runtime overhead}

We expect the runtime of practical implementations of symatch decoders to scale quite favorably. Currently however, we have not performed this runtime overhead analysis for the following reasons: Firstly, our laptop implementations of symatch and simplex-symatch run serially, i.e. $\Sigma_1$ is matched before $\Sigma_2$, before $\Sigma_{1+2}$ and so on. In practice, all the matching subroutines should be run in parallel. Additionally, simplex-symatch currently uses brute force decoding of the outer classical simplex code. While this is still tractable as BB codes have small $K$, this incurs a significant runtime cost, as it uses a number of low-weight solutions scaling exponentially in $K$.

Further, apart from the backend of \texttt{PyMatching}~\cite{pymatchingv2,gidney2021stim} for individual matching subroutines, all of our code is implemented purely in Python~\cite{symatchgit}. In comparison, BP-OSD and tesseract have a faster C++ backend.
These factors inflate our current runtime compared to a desired practical implementation. In future work, it would be useful to construct an efficient parallelized implementation and compare it against state-of-the-art decoders.

\section{Conclusion}\label{Sec:Conc}

In this work, we have proposed an efficient matching-based decoding algorithm for two-dimensional topological stabilizer codes based on symmetries. We implement our decoder on several examples of bivariate bicycle codes. 
Our decoding strategy demonstrates logical error rates that are competitive with state-of-the-art alternatives. 
Importantly, our decoder admits efficient practical implementation by circumventing time-intensive steps such as the ordered-statistics decoding of BP-OSD.

Our decoder natively performs well for a subclass of BB codes with $d_\Sigma = d$. In general, its performance is improved significantly by combining our matching subroutine with belief propagation.  
One can envision pipelining our matching step to supplement other decoding subroutines,
for instance combining our efficient matching decoder with relay BP~\cite{muller2025improvedbeliefpropagationsufficient} or vibe-BP~\cite{koutsioumpas2025colourcodesreachsurface} for shots that do not converge.
In future work, it will also be interesting to find alternative deterministic methods of passing information between symmetries that improve performance.

Our investigation also motivates consideration of how a matching-based decoder might generalize to other codes. 
Let us first remark that our decoding strategy can generalize to topological stabilizer codes that do not necessarily respect translational invariance. Indeed, provided the lattice is sufficiently large, and we are given an overcomplete check matrix that enables extraction of code symmetries, we can use the cylinder trick exactly as we have described to find a correction.
On the other hand, is as yet unclear how one can use code symmetries to produce a decoder in general. 
A key innovation we use is the cylinder trick, where we reinterpret the bivariate bicycle code as a topological code and assume its logical operators must follow a similar structure. This intuition follows from the mapping between general stabilizer codes defined on a two-dimensional lattice and copies of the toric code. 
It may be interesting to see if this intuition can be extended further to develop matching decoders for other LDPC QECCs.

We can also readily generalize our decoder to deal with correlated errors, such as correlations introduced by circuit-level noise, by modifying the symmetry graphs to account for these correlations in the conventional way. 
Further, we may adapt our decoder to correct errors in other components of a fault-tolerant quantum-computing architecture, such as generalized lattice surgery in a bivariate-bicycle code architecture~\cite{Yoder2025tour,cohen2022low,cross2025improvedqldpcsurgerylogical,williamson2024lowoverheadfaulttolerantquantumcomputation,swaroop_2025universal,he2025extractorsqldpcarchitecturesefficient}. While this is a nontrivial generalization, we suggest that it would be possible to repeat our process of designing a decoder by searching for symmetries of the merged spacetime code that combines the logical code blocks together with a logical processing unit. 
It is interesting to ask how logic gates on BB codes map to gates on the equivalent symmetry-enriched toric codes, and how this can be used to extend our decoders to this setting.

Finally, let us briefly remark that it has been enlightening to concentrate on a simple noise model that enables us to benchmark decoder performance by exhaustively interrogating low weight errors. 
At the beginning of the present endeavor, it was our goal to find a decoder that could correct all errors of weight $w <d/2$. We indeed achieve this goal for a subset of BB codes. However, we fall short on this metric for the gross code, as even our best performing decoder fails on  $\sim 500$ of the 48-million weight $5 < d/2 = 6$ errors.  
We expect that this might improve using more sophisticated methods of passing information between matching subroutines. 
We hope our work encourages research finding a decoder that can correct up to weight $d/2$ errors for code capacity noise. Indeed, one expects the fault-tolerant generalization of any such decoder to be able to correct all low-weight errors.

\textit{Acknowledgements-- }
We are grateful to A.~Cross, T.~Jochym-O'Connor, S. Puri, S. J. S. Tan, and T.~Yoder for insightful discussions. 
We thank H. Westerheim and T. B. Smith for comments on an early version of the manuscript. 
BJB thanks the Center for Quantum Devices at the University of Copenhagen for their hospitality. Code for symatch decoders is available at~\cite{symatchgit}.

\textit{Note added--} After the completion of this manuscript, we became aware of upcoming related work on generalizing matching decoders to 2DTI codes~\cite{tan2026generalizedmatchingdecoders2d}.

\bibliography{LDPCcodes}

\bibstyle{plain}

\appendix

\section{Bivariate bicycle codes in the check-matrix formalism} \label{App:CheckMatrix}

Here, as an alternative to the polynomial formalism used in \cref{Sec:Prelim}, we present bivariate bicycle codes in terms of the check matrix formalism.
We define quantum error-correcting codes in terms of their stabilizer group. 
Stabilizers $\sigma\in \mathcal{S}$ generate an abelian subgroup of the Pauli group such that $-1\not\in \mathcal{S}$. 
Logical states $|\psi\rangle$ protected by the code satisfy $\sigma |\psi\rangle = (+1) |\psi\rangle$ for all elements $\sigma\in\mathcal{S}$. 

We express a generating set of stabilizers using the check matrix $H$ of a code. This is a $2n \times m $ binary matrix where $ n $ is the number of qubits of the code and $ m \ge n - k $ where $k$ is the number of encoded logical qubits. 
Each row of the check matrix $\vec{h}$ defines a stabilizer generator. The elements of $\vec{h} = \left( \left. \vec{q} \  \right|  \vec{t} \ \right)$ can be divided into two $1 \times n$ row vectors $\vec{q}$ and $\vec{t}$ such that stabilizers are of the form
\begin{equation}
\sigma \left( \left. \vec{q} \  \right|  \vec{t} \ \right) = \prod_{j = 1}^n  Z_j^{q_j} X_j^{t_j}, \label{Eqn:PauliAsVector}
\end{equation}
where $X_j$ and $Z_j$ denotes the standard Pauli matrix acting on the $j$-th qubit of the system and $q_j$ ($t_j$) denote the $j$-th element of $\vec{q}$ ($\vec{t}$).

We can write check matrices of a BB code~\cite{Kovalev2013fault, Bravyi2024high} in terms of $\ell \times \ell$ matrices of cyclic shifts $C_\ell$. Let us write down a canonical basis of vectors $\vec{e}_j$ whose $j$-th element is $1$ and all other elements are $0$, such that $C_\ell \vec{e}_j = \vec{e}_{j+1} $ for all $j$ with periodic boundary conditions such that $\ell+1 = 1$. We can define cyclic shifts about higher dimensional lattices using a tensor product structure. For instance we have a periodic two-dimensional lattice of dimensions $M \times N$ such that
\begin{equation}
x = C_M \otimes I_N, \quad y = I_M \otimes C_N, \label{Eqn:TranslationOperators}
\end{equation}
where $I_\ell$ is the $\ell \times \ell$ identity matrix. We note that we have deliberately used the same terms $x$ and $y$ for these cyclic shifts as we have used for terms in polynomial equations in the main text.

Bivariate bicycle codes are Calderbank-Shor Steane codes~\cite{Calderbank1996good, Steane1996error}. This means we can express the check matrix in a block diagonal form
\begin{equation}
H = \begin{pmatrix}
H_X & 0 \\ 0 & H_Z
\end{pmatrix}.
\end{equation}
where $H_X$ ($H_Z$) describes stabilizers that are the product of Pauli-$X$ (Pauli-$Z$) operators only. These matrices must satisfy $H_X H_Z^T = 0$ so that all stabilizers commute.

We express $H_X$ and $H_Z$ for bivariate bicycle codes in terms of $MN \times MN$ matrices $A$ and $B$.
The matrices $A$ and $B$ are representatives of the polynomials presented in the main text. We then write down the $ MN \times 2MN $ block matrices of the check matrix as follows: 
\begin{equation}
H_X = \left[ \left. B^T \right| A^T\right], \quad H_Z = [A | B]. 
\end{equation}
We note that the first $MN$ columns of these two matrices denote the support of the stabilizer on the left qubits of the lattice, and the second $MN$ columns denote the support of the stabilizer on the right qubits.

In order for the stabilizer group to commute, i.e., $H_X H_Z^T = 0$, it follows that $A$ and $B$ must commute. We note that there is a natural division of the qubits of $H_X$ and $H_Z$ in terms of the first $MN$ elements of a row and the second $MN$ elements of each row. These are commonly referred to as the left qubits and right qubits, respectively. 
Throughout the manuscript we represent the code with qubits on the edges of a square lattice where the left (right) qubits are supported on horizontal (vertical) lattice edges, see Fig.~\ref{Fig:stabilizers}.

\subsection{Error syndrome and the noise model}
\label{SubSec:GaussianElimination}

We can write a Pauli error as a $2n \times 1$ column vector $\vec{e} = \left( \vec{e}_Z | \vec{e}_X \right)^T $ with $\vec{e}_X$ ($\vec{e}_Z$) an $1 \times n$ vector denoting the Pauli-$X$ (Pauli-$Z$) component of the error. Specifically, we have that $\vec{e}_X$ ($\vec{e}_Z$) takes value 1 on element $j$ to denote a Pauli-$X$ (Pauli-$Z$) error on qubit $j$ and otherwise the element takes value $0$. We use the check matrix to return an error syndrome $\vec{s}$. The error syndrome is a $m \times 1$ vector such that
\begin{equation}
\vec{s} = H \vec{e}.
\end{equation}
Elements of the error syndrome report which stabilizers are violated by a given error. An element $s_j$ of $\vec{s}$ takes value $1$ if the stabilizer generator denoted by the $j$-th row of $H$ is violated by error $\vec{e}$. Its value is $0$ otherwise.

\begin{figure*}[t]
    \centering
    \includegraphics[width=1\linewidth]{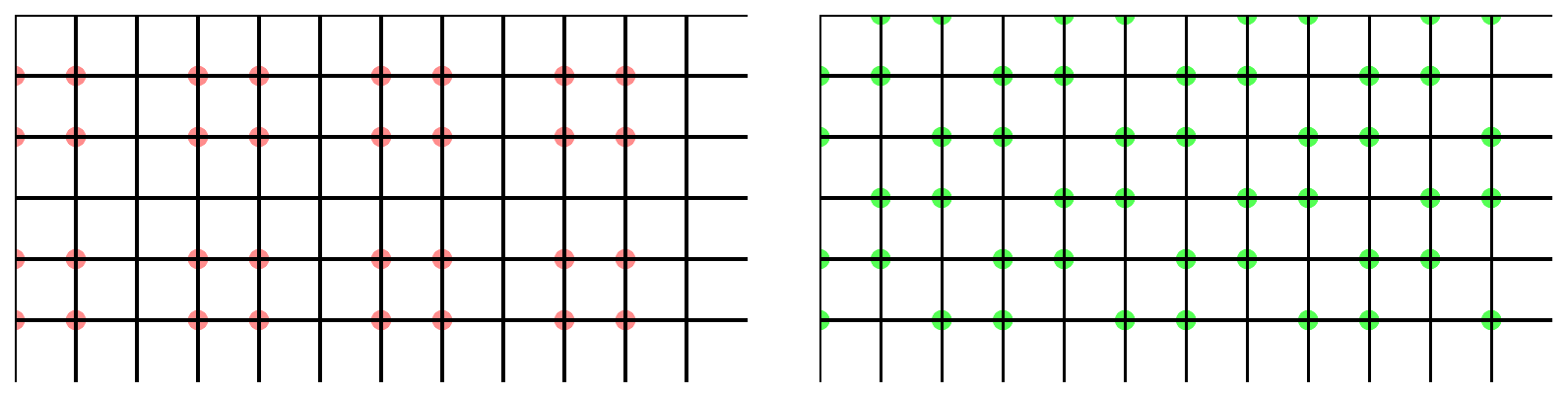}
    \caption{A 32-check and 48-check symmetry of the gross code. Note the gridlike structure of the left symmetry, and the paired diagonal rows of the right symmetry. These respectively mirror symmetries of HGP codes and color codes.}
    \label{fig:syms3248}
\end{figure*}

\subsection{Finding symmetries with the check-matrix formalism}

The check matrix formalism gives us a straightforward way of finding BB code symmetries. This is achieved using Gaussian elimination on the over-complete check matrix. Let us write down the expression
\begin{equation}
\mathtt{RR}(H) = \Gamma H,
\end{equation}
where $\mathtt{RR}(\cdot)$ denotes the row-reduced echelon form of its argument and $\Gamma$ denotes the row operations needed to achieve complete row-reduction on check matrix $H$. Given that $H$ is over-complete, this operation gives rise to a matrix with $m - n + k \ge 0 $ rows of all zero elements. In the check-matrix formalism, these rows with all null elements have corresponding symmetries that can be read off from the matrix $\Gamma$. Let us verify this result in the remainder of this subsection.

Recall a symmetry is some set of over-complete stabilizer generators $\Sigma$. Let us denote a standard set of stabilizer generators $\sigma_j$ with $1 \le j \le m$ where now $\sigma_j$ corresponds to the $j$-th row of $H$. We therefore look for sets of stabilizer generators $\Sigma$ such that
\begin{equation}
\prod_{\Sigma} \sigma_j = 1. 
\end{equation}

We have that 
\begin{equation}
M := \mathtt{RR}(H) =  \sum_\ell \Gamma_{j,\ell} H_{\ell, k},
\end{equation}
where rows of $M$ describe elements of the stabilizer group where the rows express Pauli operators as defined in Eqn.~(\ref{Eqn:PauliAsVector}).
Equivalently, we have a new set of stabilizer generators $\sigma'_j$ expressed as follows
\begin{equation}
\sigma'_j = \prod_\ell \sigma_\ell^{\Gamma _{j,\ell}}.
\end{equation}
where $\sigma_\ell$ is the stabilizer corresponding to the $\ell$-th row of $H$ and $\sigma'_\ell$ corresponds to the stabilizer generator at the $\ell$-th row of $ M$.
 We look for elements $\sigma'_j= 1$ or equivalently rows of $M$ such that
\begin{equation}
M_{\ell, k} = 0, \quad \forall k.
\end{equation}
Given an over-complete check matrix $H$ and $\Gamma$ such that $M = \mathtt{RR}(H)$ has rows with non-zero elements we can read off the symmetries from $\Gamma$. For each row $j$ of $\mathtt{RR}(H)$ with all zero elements, we have the $j$-th row of $\Gamma$ such that
\begin{equation}
\prod_\ell \sigma_\ell^{\Gamma _{j,\ell}} = 1.
\end{equation}
The elements of the rows where $\Gamma_{j,\ell} = 1$ then correspond to $\Sigma$, the set of elements of the initial generating set of stabilizers $\sigma_j$, i.e., the rows of $H$, that give rise to a symmetry.

\section{Examples of code symmetries} \label{App:SymPics}

Here, we show some additional symmetries of the BB144 code. \cref{eq:transSym} states that a generating symmetry set may be obtained from translations of the 36-check symmetry shown in \cref{Fig:Symmetry}. Interestingly, linear combinations of this generating set produce 32-check symmetries which mirror symmetry structures observed in HGP codes, and 48-check symmetries which mirror the symmetry structure of color codes~\cite{Delfosse_2014decoding}. These symmetries are drawn in ~\cref{fig:syms3248}, and are used during simplex over-matching.

\subsection{Constructing symmetries for an infinite BB code}
\label{app:InfBBCode}

We now describe a procedure to find symmetries in an infinite BB code.
For a BB code specified by 
\begin{align}
    \begin{pmatrix}
        1+g(x^{-1},y^{-1}) & 0
        \\
        1+f(x^{-1},y^{-1}) & 0
        \\
        0 &  1+f(x,y)
        \\
        0 &  1+g(x,y)
    \end{pmatrix}
\end{align}
we focus on the $Z$-checks, noting the $X$-checks are related by an equivalence. We consider a polynomial
\begin{align}
    \Sigma_f^{L} :&= (1+f+f^2+\cdots+f^{L-1})\\
     &= (1+f)(1+f^2)\cdots(1+f^{L/2})
\end{align}
for $L=2^{\ell}$.
We have that 
\begin{align}
    \Sigma_f^{L_f}\Sigma_g^{L_g}
        \begin{pmatrix}
         1+f
        \\
         1+g
    \end{pmatrix}
    =
    \begin{pmatrix}
         (1+f^{L_f}) \Sigma_g^{L_g}
        \\
         (1+g^{L_g}) \Sigma_f^{L_f}
    \end{pmatrix}.
\end{align}
Using the above, we can find $Z$-check symmetries on any $L_x \times L_y$ torus, i.e.~$x^{L_x}=1,y^{L_y}=1$, where ${f^{L_f}=1},{g^{L_g}=1},$ for some $L_f=2^{\ell_f},L_g=2^{\ell_g}$.
In particular, we have the following symmetries
\begin{align}
    x^iy^j\Sigma_f^{L_f}\Sigma_g^{L_g} ,
\end{align}
for all $i,j$. 
These symmetries can then be tiled over $L_x \times L_y$ sized regions to cover the infinite plane.

\section{Toric code equivalence and topological defect networks}
\label{app:TDN}

In this appendix we discuss the mapping of translation-invariant two-dimensional stabilizer codes to copies of the toric code from a defect topological quantum field theory perspective.

Any family of 2DTI stabilizer codes is equivalent to a number of copies $K$ of the toric code enriched by a $\mathbb{Z}\times \mathbb{Z}$ translation symmetry action~\cite{Bombin2012universal,Bombin2014structure,haah2016algebraic,Haah2021classification}. 
There is a tension between the constant number of logical qubits encoded into copies of the toric code, and the more complicated functional dependence of the number of logical qubits on the system size for a general 2DTI quantum LDPC code. 
This is resolved by considering 2D translation-symmetry-enrichment of the toric codes, which is realized in the general 2DTI code~\cite{Dua2019}. 
The $\mathbb{Z}\times \mathbb{Z}$ translation group may act nontrivially on the anyon types. 
This action factors through a finite quotient group $\mathbb{Z}_{R_x}\times \mathbb{Z}_{R_y}$, as the number of anyon types and the size of their automorphism group are both finite. 
In particular, any 2DTI code with a growing distance can be realized via compactification of a higher dimensional translation-invariant fracton code with no stringlike logical operators along the compactified directions~\cite{Hopkin2025}. 
The BB codes in particular are compactified hypergraph-product fracton models~\cite{Tan2023}. 

\begin{figure}[t]
    \centering
    \includegraphics[page=1]{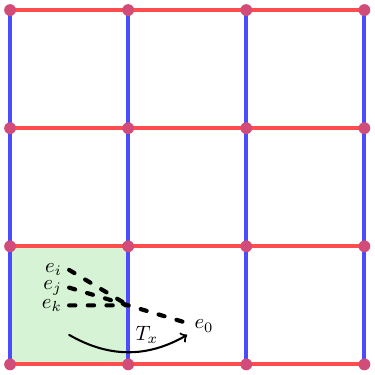}
    \caption{The topological defect network representation of a 2DTI code. The green square denotes a unit cell containing copies of the toric code. 
    The blue and red lines denote invertible topological defects that may permute anyons between the copies of toric code. 
    $T_x$ denotes a translation which moves an anyon $e_0$ one unit along the $x$-axis. 
    This results in a new anyon type in the fundamental (green) unit cell obtained via applying a string operator to move the $e_0$ anyon through the blue domain wall, resulting in a product of anyons over several layers $e_i e_j e_k$. }
    \label{fig:TDN}
\end{figure}

When the nontrivial translation action is preserved, the equivalence to copies of toric code becomes enriched by symmetry. 
This results in a topological defect network~\cite{Else2018,Aasen2020,Song2023} (TDN) given by copies of toric code, coupled together by nontrivial invertible domain wall defects along the boundaries of each unit cell, see Fig.~\ref{fig:TDN}. 
A similar construction was used in Ref.~\cite{Brown2013Entropic} for the purposes of increasing the error-correcting performance of the qudit toric code.
The central feature of the topological defect network representation of a 2DTI code is that all topological excitations can be represented in the fundamental unit cell, and the action of translation along the $x$-axis ($y$-axis) is equivalent to moving the translated anyon through the vertical (horizontal) domain wall, returning it to the fundamental unit cell. 
For CSS 2DTI codes, this can only result in a permutation between $e$ anyons ($m$ anyons) that preserves the exchange statistics. 

The presence of the domain walls determines the behavior of the ground state degeneracy as a function of the system size (in terms of unit cells). 
The TDN on a torus of size $L_x \times L_y$ is equivalent to copies of the toric code with a single domain wall along the $x$-axis ($y$-axis) given by the fusion product of all $x$-oriented ($y$-oriented) domain walls in the TDN. 
The degeneracy of this codepsace is known to be equal to the number of $T_x^{L_x}$-invariant $T_y^{L_y}$ defects~\cite{barkeshli2014symmetry}. 
For copies of the toric code, this is equal to the number of anyons that are invariant under both $T_x^{L_x}$ and $T_y^{L_y}$. 
This can be understood via the conditions that the logical string operators for anyons $a$ in the $x$-direction must satisfy $T_x^{L_x} a = a$, and there is an additional equivalence relation given by $T_y^{L_y} a\sim a$, similar for the $y$-direction. 
This provides a purely 2D topological explanation for the complicated code space degeneracy as a function of system size. 
It also explains why there is some length $R_x$ ($R_y$), given by the order of the symmetry action on the anyons $T_x^{R_x}=\mathds{1}$, after which the pattern of ground space degeneracies repeats. 
For an unfrustrated $R_x \times R_y$ torus, there are effectively untwisted boundary conditions, resulting in $2K$ logical qubits which is the maximum number of logical qubits achievable for any system size.

The local unitary equivalence between a general 2DTI code and copies of the toric code may require a circuit of a large constant depth. 
It is only at the scale of this circuit depth that properties such as deformable stringlike logical operators become apparent. 
Despite this limitation, some aspects of the effective toric code picture can be applied directly to the original 2DTI code without performing a local unitary. 
The symmetries and logical operators on the unfrustrated torus can be matched to the anyons of the effective toric code description. 
On frustrated tori, the symmetries can then be identified with the local symmetries of multiple anyon types, that are related by passing through the twisted boundary conditions. 
This information is relevant when finding a correction operator for a neutral syndrome cluster, as clusters satisfying all local symmetries are locally correctable, while more general neutral clusters may require a correction that passes through the boundary condition that relates the local symmetries for multiple anyon types. 
The importance of considering the boundary condition is captured by having an appropriate distance function on the matching graph. 
For example in two copies of toric code on a frustrated torus that includes a layer swap defect around one handle, a pair of adjacent $e_0,e_1,$ anyons is globally neutral, but requires a potentially high weight correction operator that passes through the layer swap defect.

Finally, the toric code picture also reveals the simple behavior of the number of logical qubits in a 2DTI code with surface code type boundary conditions of truncated $X$-checks on vertical boundaries, and truncated $Z$-checks on horizontal boundaries (or vice-versa)~\cite{Aitchison2024}. 
This simply corresponds to $e$-condensing horizontal boundaries and $m$-condensing vertical boundaries on the TDN. 
The invertible domain walls can then be fused into these boundaries, and are absorbed by them. 
This results in an equivalence to the standard decoupled patches of surface code with $e$-condensing horizontal boundaries and $m$-condensing vertical boundaries.

\section{Calculating equivalence to toric codes}
\label{app:TCE}

In this appendix we discuss an approach to calculating the number of toric codes that are equivalent to a 2DTI code and their translation-symmetry-enrichment following Ref.~\cite{Dua2019}. 

\begin{figure}[t]
    \centering
    \includegraphics[page=2]{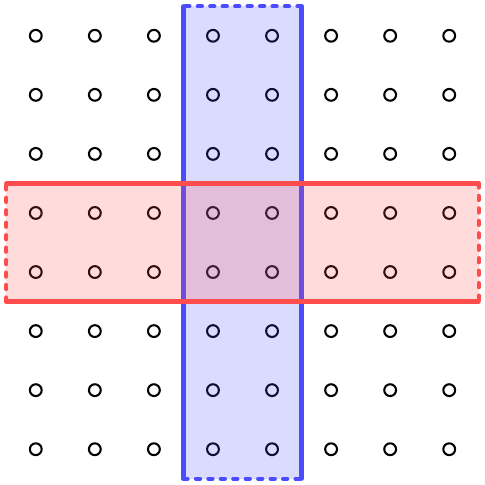}
    \caption{The regions used to find lattice string operators along the $x$-axis (red) and $y$-axis (blue). 
    The solid lines denote closed boundary conditions, where all overlapping checks are enforced. 
    The dashed lines denote open boundary conditions, where checks that act both inside and outside the region are not enforced. 
    This allows string operator segments to end on the open boundaries.
    The restriction of these string operators to the overlapping region can be used to compute their commutation relations.
    }
    \label{fig:LatticeS}
\end{figure}

To determine the number of toric codes equivalent to a given 2DTI code, the topological entanglement entropy provides a simple method. However, the result of this calculation may not be reliable due to spurious contributions~\cite{PhysRevB.94.075151,Williamson2018}. 
A reliable method is based on the lattice $S$-matrix invariant~\cite{haah2016Invariant,Dua2019}. 
This method considers a pair of crossing horizontal and vertical strips on a plane of the 2DTI code. 
These strips can be finite, but must be sufficiently long and sufficiently wide, which can be ensured in practice by increasing the length and width until a stable result is obtained. 
The boundary conditions on the strips allow excitations or syndromes at the endpoints only, see Fig.~\ref{fig:LatticeS}. 
A basis of string operator segments can be found by diagonalizing the commutation matrix between the string operators supported on the horizontal and vertical strips, respectively. 

The action of the $T_x$ ($T_y$) translation symmetry on the anyons can be computed by shifting the vertical string segments by $T_x$ and using the braiding statistics with the unshifted horizontal strings to resolve the resulting string types using braiding modularity.

Below, we describe in detail the calculation of the string operator segments for a 2DTI code. Next, we describe the translation action on these string segments, which determines the size of the unfrustrated unit cell or tori. 
We then discuss how to find a basis for the full set of unfrustrated symmetries and the associated string operators. 

\subsection{String operator segments}

We first calculate a reliable invariant that characterizes the number of copies of toric code and the action of the 2D translation symmetry upon them. 
For this, we consider a method based on the lattice version of the $S$ matrix invariant for anyon theories. 
This method first constructs a set of horizontal and vertical string operators on fattened ribbons, as shown in Fig.~\ref{fig:LatticeS}. 
These string operators can be found by taking the kernel of a restricted check matrix that keeps all checks with support on the ribbon, except those whose support on the ribbon is confined to the end strips, which are highlighted in blue in Fig.~\ref{fig:LatticeS}. 
The kernel of the restricted horizontal check matrix defines a large set of string operators, 
$$\{s^{H}_{i}\}_i,$$ 
which includes nontrivial string segments, and stabilizers. 
Similarly, we have 
$$\{s^{V}_{j}\}_j,$$ 
for the vertical string segments. 
For simplicity we focus on CSS codes and assume the horizontal string operators are $X$-type and the vertical string operators are $Z$-type. Each element $s^{H}_{i},s^{V}_{j},$ then corresponds to a binary column vector representative of the associated Pauli string operator. 
The analogous calculation with the direction of the $X$ and $Z$-type string operators reversed proceeds similarly. 
We collect the horizontal string operator column vectors into a matrix $s^{H}$ and similarly for the vertical string operators $s^V$. 
To find a basis of nontrivial string segments, we construct the commutation matrix
$$C=(s^H)^T (s^V),$$
which has matrix elements  
$$C_{i,j}=s^{H}_{i} \cdot s^{V}_{j}.$$ 
We diagonalize $C$ to find a basis of anticommuting string segments. 
This can be achieved via the binary Smith normal form
$$C=U D W,$$
where $U$ and $W$ are invertible binary change of basis matrices, and $D$ is a diagonal matrix with an identity block of length equal to the number of independent anticommuting string segment pairs $K$, direct summed with a zero block. The number $K$ here yields the number of independent copies of toric code in the model~\cite{Dua2019}. 

The above decomposition lets us pick out a basis of $K$ nontrivial string segments from linear combinations of $s^{H}_{i}$, $s^{V}_{j}$, and throw away the remaining stabilizer operators. 
This basis of string operators is given by the first $n$ columns of the matrices as follows 
$$\hat{s}^H=(s^H)(U^{-1})^T|_{K\times K} \quad \text{and}\quad \hat{s}^V=(s^V)W^{-1}|_{K\times K} .$$
The number of toric code copies $K$ is an invariant under finite depth, finite range quantum circuits. 

For this procedure to produce a stable and correct answer, the ribbons must be sufficiently wide, with the length much larger than the width, and the end strips must be sufficiently thick. 

\subsection{Translation action}

The next important piece of information is how the 2D translation symmetry acts on the basis of string operators. 
To calculate this, we make use of the modularity of anyon braiding. 
This means the $S$ matrix is invertible, and hence knowing the commutation relations of a string operator with a basis of string operators uniquely specifies which anyon it corresponds to. 
For our purposes, we can compute the anyon permutation action of the translation along the horizontal direction by shifting the vertical string operators by a single lattice site in the horizontal direction to construct a new matrix we call $\mathcal{T}_x\hat{s}^V$. 
We then construct the shifted commutation matrix
$$C_x=(\hat{s}^H)^T (\mathcal{T}_x\hat{s}^V),$$
and diagonalize it to find the anyon permutation action $P_x$ caused by the translation 
$$C_x= D P_x^{-1} .$$
Note that $P_x$ is not, itself, a permutation matrix but rather a change of basis on the string operator segments which determines a permutation of the anyon types. 
The action of translation along the vertical direction $P_y$ can be computed similarly by shifting the horizontal string operators. 

The translation action matrices $P_x,~P_y,$ effectively describe domain walls that sit on the left and bottom boundaries of each unit cell. 
This means that a torus with dimensions $L_x \times L_y$ is equivalent to $n$ copies of the toric code with twisted boundary conditions corresponding to $P_x^{L_x},~P_y^{L_y},$ for anyons that travel around the horizontal, vertical, cycle, respectively. 
Hence, the orders $R_x,~R_y,$ of the translation actions $P_x,~P_y,$ give the width and height, respectively, of the minimal unfrustrated unit cell which corresponds to an unfrustrated torus upon taking periodic boundary conditions.

\subsection{Unfrustrated symmetries and strings}
\label{app:Unfrustrated}

For simplicity, we focus on 2DTI codes with an equal number of checks and qubits per unit cell. 
This implies that on any finite sized torus, the number of logical qubits is equal to the number of independent symmetries. 
Further assuming that the code has a growing distance with system size, there can be no local symmetries which are fully contained within a constant sized region. Rather, all symmetries must be global. 

For the 2DTI codes we consider, on the infinite plane there are $2K$ independent symmetries, two per copy of toric code, one per generating anyon. 
Similarly, there are $2K$ symmetries on the unfrustrated torus, as it is equivalent to $K$ copies of the toric code with untwisted boundary conditions. 
Due to the periodic boundary conditions, the symmetries on the unfrustrated torus can be tiled to cover the infinite plane. 
In this way, all symmetries on the infinite plane are generated by those on the unfrustrated torus. 

The unfrustrated symmetries define a basis of fundamental string operator segments for the effective toric code anyons, which move an anyon by $R_x,~R_y$ in the horizontal, vertical, direction respectively. 
To find these string segments we use the \textit{cylinder trick}. 
For example, for the horizontal string operator we consider an infinite cylinder given by an infinitely tall vertical strip of width $R_x$ with periodic boundary conditions. 
We form a truncated symmetry by taking a symmetry on the unfrustrated unit cell, and tiling it onto the upper half strip. 
This leaves a dangling operator wrapping the cylinder on a ribbon around the position $y=0$. 
This operator is guaranteed to be a logical operator as it anticommutes with any string operator segment that creates a pair of anyons involved in the symmetry, sufficiently far above and below $y=0$.
These anyons can then be pushed off to $y\rightarrow \pm \infty$ to find an anticommuting logical operator partner. 
A similar strategy applies to find vertical fundamental string operator segments. 
We remark that the above argument can similarly be applied to show that string operators on frustrated tori obtained from the cylinder trick are nontrivial. 

By using the fundamental string segments, we can find a basis for the symmetries that precisely corresponds to decoupled toric codes. 
This is done by diagonalizing a commutation matrix between the horizontal and vertical string segments, and using the change of basis matrix found there on the symmetries. 
At this point, we have a basis of symmetries and associated string operators on any torus given by tiling unfrustrated unit cells with periodic boundary conditions. 

On frustrated tori, the symmetries take the form of a product of the unfrustrated symmetries on local patches. However, these symmetries may be joined together by the frustrated boundary conditions. This is accounted for by assigning appropriate weights to the matching graphs on these frustrated symmetries that reflect the weight of the shortest operator that creates a given pair of syndromes in the symmetry. This weight may be much larger than the separation between the syndromes in the tanner graph, as it may require an operator that wraps nontrivially around a handle of the torus. We note that appropriate weights are assigned automatically for a matching graph that is constructed by connecting the checks contained in a symmetry via local errors that anticommute with them.

\section{Correcting matching errors with a simplex encoding}
\label{App:Hamming}

With the cylinder trick, matching ultimately returns a bit value that corresponds to the commutator between a logical operator and the error the code has experienced. We might expect a matching result to be unreliable. As described in \cref{SubSec:Simplex}, we therefore look for consistency conditions among matching results to identify errors in matching results. We find that the bits returned from an overcomplete set of matching results can be encoded in a simplex code. Here, we describe the checks for the simplex code where we have $K$ logical operators about a non-trivial cycle, and we explain how to exhaustively search for a correction.

\subsection{Checks from constraints among matching results}

In order to define checks among matching results let us begin by establishing notation. We first define a set of $K$ independent logical operators $\overline{Z}_1$, $\overline{Z}_2$,\dots, $\overline{Z}_K$ that wrap around the same non-contractible cycle. These logical operators are found using the cylinder trick on $K$ independent symmetries $\Sigma_1$, $\Sigma_2$,\dots, $\Sigma_K$, respectively. It will also be helpful to define a vector $\vec{v} = (v_1, v_2, \dots , v_K)$ of binary values $v_j = 0,1$ such that we can define general symmetries of the symmetry group in terms fo these vectors as follows:
\begin{equation}
\Sigma[\vec{v}] := \sum_{j=1}^K v_j \Sigma_j,
\end{equation}
such that each vector defines linear combination of the generating set of symmetries.

The vector notation also helps us to define general logical operators. Using the cylinder trick for symmetry $\Sigma[\vec{v}]$ we obtain the logical operator
\begin{equation}
\overline{L}[\vec{v}] = \prod_{j = 1}^K \overline{Z}_j^{v_j}.
\end{equation}
Let us also define the matching result $b[\vec{v}]$ as the bit value returned from matching on symmetry $ \Sigma[\vec{v}] $ as the estimated number of bit flips supported on logical operator $\overline{L}[\vec{v}] $. We assume that a matching result may be unreliable. We therefore treat these variables $b[\vec{v}]$ as bits that may have experienced an error. It is our goal to determine which bits have experienced an error. 

In order to identify errors in $b[\vec{v}]$, we look for sets of distinct vectors $\vec{v} \in \mathcal{V}$ such that the sum of bits $b[\vec{v}]$ are constrained to give a constant value, i.e.
\begin{equation}
\sum_{\vec{v}\in\mathcal{V}_C} b[\vec{v}] = \textrm{constant}, \label{Eqn:GenericCheck}
\end{equation}
where the summation is made modulo 2. These sets of vectors $\mathcal{V}$ therefore give rise to checks $C$ that can identify errors on our bit values $b[\vec{v}]$ assuming we obtain matching results that violate constraints as in Eqn.~(\ref{Eqn:GenericCheck}).

We find constrained sets of vectors $\mathcal{V}$ by looking at relations among logical operators. We begin with a set of logical operators found using the cylinder trick that satisfy the following relation
\begin{equation}
\prod_{\vec{v}\in\mathcal{V}}\overline{L}[\vec{v}] = 1, \label{Eqn:LogicalConstraint}
\end{equation}
where all logical operators are obtained using the same cylinder over the same non-contractible cycle of the torus. The relationship among logical operators written above implies the following relationship among bits returned from matching results
\begin{equation}
\sum_{\vec{v}\in\mathcal{V}} b[\vec{v}] = 0,
\label{Eqn:FullSupportCheck}
\end{equation}
where the equation above is obtained by recursively applying the following identity to Eqn.~(\ref{Eqn:LogicalConstraint}) 
\begin{equation}
\overline{L}[\vec{v}_1 + \vec{v}_2] = \overline{L}[\vec{v}_1] \overline{L}[\vec{v}_2], 
\end{equation}
and using that $\overline{L}[\vec{0}] = 1$ where $\vec{0}$ is the null vector.

We therefore reduce the problem of finding constrained sets of bits to finding sets of vectors $\mathcal{V}$ satisfying
\begin{equation}
\sum_{\vec{v}\in \mathcal{V}} \vec{v} = \vec{0}, \label{Eqn:Constraint}
\end{equation}
with addition taken modulo 2.
Similar to the example presented in the main text then, we can define a check $C$ for each $\mathcal{V}$ such that
\begin{equation}
C := \sum_{\vec{v}\in\mathcal{V}_C} b[\vec{v}], \label{Eqn:SupportCheck}
\end{equation}
where we expect that $C=0$ in the case that all matching results are correct. An odd number of errors in the matching results $b[\vec{v}],\vec{v} \in \mathcal{V}_C$ will violate a check.

It therefore remains to find vector sets $\mathcal{V}$ satisfying Eqn.~(\ref{Eqn:Constraint}). We use two properties that are easily checked to find these constraints. First, we have $ \sum_{\vec{v} \in \mathbb{Z}_M} \vec{v} = \vec{0} $ for $M \ge 2$ and we take the sum modulo 2. Secondly, we have that there are an even number of vectors $\vec{v} \in \mathbb{Z}_M$. 
It follows immediately from the fact above that we have a global check $C$ with $\mathcal{V}$ that including all vectors $\vec{v} \not= \vec{0}$, where we assume the bit $b[\vec{0}] = 0 $ will necessarily give a trivial result since it corresponds to the result of matching over a trivial symmetry $\Sigma[\vec{0}]$.

We can find additional constraints by choosing sets of vectors with specific elements of each vector fixed to 1. For example, let us define sets of vectors as follows
\begin{equation}
\mathcal{V}[v_j = 1, v_k = 1, \dots] = \left\{ \vec{v} \in \mathbb{Z}_K : v_j = 1, v_k = 1, \dots \right\}, 
\end{equation}
where the list in square braces $v_j = 1, v_k = 1, \dots$ indicates which elements of each vector in the set are fixed to $1$. Otherwise, the set contains all possible vectors in this set for elements whose values are not constrained.

As we make frequent use of sets of vectors with constrained elements of $\vec{v}$, let us propose some shorthand notation. Let us write a vector $ \vec{c}\in \mathbb{Z}_K $ to denote a set of constraints $\{\vec{c}\}$ such that all vectors $\vec{v}\in\mathcal{V}[ \{ \vec{c} \}]$ must have $v_j = 1$ if $c_j = 1$, otherwise elements $v_j$ can take any value if $c_j = 0$. Let us also define a check in terms of constraint sets $\{\vec{c}\}$. We define the check 
\begin{equation}
C[\{\vec{c}\}] = \sum_{\vec{v}\in\mathcal{V}[\{\vec{c}\}]} b[\vec{v}]. 
\end{equation}
The check defined above is the special case with $C[\{\vec{c} = \vec{0}\}]$. Let us now look for other value choices of constraint sets $\{\vec{c}\}$.

In general we find that we can fix up to $K-2$ elements of each vector $\vec{v}$, i.e., $|\vec{c}| \le K-2$ where $|\cdot|$ denotes the hamming weight of a vector. Explicitly, we have checks
\begin{equation}
C  [\{ \vec{c} \}] =  \sum_{\vec{v}\in  \mathcal{V} [\{ \vec{c} \}] } \vec{v},
\end{equation}
for all $|\vec{c}| \le K-2$.
is satisfied. To verify this we need only check that
\begin{equation}
\sum_{\vec{v}\in\mathcal{V}[\{\vec{c}\}]} \vec{v} = \vec{0}, \label{Eqn:CheckCondition}
\end{equation}
for all $|\vec{c}| \le K-2 $.

We can easily see this equation holds by expressing all vectors $\vec{v}\in\mathcal{V}[\{\vec{c}\}]$ in terms of their free and fixed terms 
\begin{equation}
\vec{v} = \vec{v}_\textrm{fixed} \oplus \vec{0}_\textrm{free} + \vec{0}_\textrm{fixed}\oplus \vec{v}_\textrm{free}
\end{equation}
as determined by the constraint set $\{\vec{c}\}$ where $\vec{v}_\textrm{fixed} = \vec{1}$ are the $|\vec{c}| \le K-2$ elements of $\vec{v}$ fixed to $1$ due to constraints $\{\vec{c}\}$, and $\vec{v}_\textrm{free}\in\mathbb{Z}_M$ are the remaining $M = K - |\vec{c}| \ge 2 $ elements of $\vec{v}\in\mathcal{V}[\{\vec{c}\}]$ that can take free values. We are therefore essentially evaluating
\begin{equation}
\sum_{\vec{v}\in\mathcal{V}[\{\vec{c}\}]} \vec{v} = \sum_{\vec{v}\in\mathcal{V}[\{\vec{c}\}]} \left( \vec{v}_\textrm{fixed} \oplus \vec{0}_\textrm{free} + \vec{0}_\textrm{fixed}\oplus \vec{v}_\textrm{free} \right), 
\end{equation}
where we consider the two parts of the sum separately.

Let us deal with the sum over the free terms first; $\sum \vec{0}_\textrm{fixed}\oplus \vec{v}_\textrm{free}$.
Using that in the non-trivial part of this sum, we take the sum over all free vectors $ \vec{v}_\textrm{free}\in\mathbb{Z}_M $ and $M \ge 2$, we have that all the free terms vanish over the summation $\sum_{\vec{v}_\textrm{free}\in\mathbb{Z}_M} \vec{v}_\textrm{free} = \vec{0} $. 
Then we consider the part of the summation $\sum \vec{v}_\textrm{fixed} \oplus \vec{0}_\textrm{free}$. We are only interested in the case where $\vec{c}\not=\vec{0}$ as otherwise the fixed term is trivial. Given that we sum over an even number of vectors since there are an even number of free terms, i.e., $|\mathcal{V}[\{\vec{c}\}| = 2^M$, we have that $\sum \vec{v}_\textrm{fixed} \oplus \vec{0}_\textrm{free} = 2^M \vec{1} \oplus \vec{0} = \vec{0} \oplus \vec{0}$ given that we are taking summations modulo 2 and $2^M$ is even. All together these results verify Eqn.~(\ref{Eqn:CheckCondition}).

Finally, it is helpful to note that with the exception of $C[\{\vec{0}\}]$, all checks include an even number of bits since we have $2^M$ free combinations for vectors $\vec{v}\in\mathcal{V}[\{\vec{c}\}\}]$. As mentioned, $C[\{\vec{0}\}]$ has support on an odd number of bits since we exclude the bit $b[\vec{0}] = 0$ as trivial.

\subsection{Correcting the simplex code}

The code we have described is a simplex code with parameters $[ 2^K - 1, K, 2^{K-1} ]$. Let us briefly describe how we can find a least weight correction by exhaustive search. At a high level, we find an initial correction by modifying bit values $b[\vec{v} = \vec{c}]$ for each check $C[\{\vec{c}\}]$ where we begin with all checks $C[\{\vec{c}\}]$ with the largest constraint weight $|\vec{c}| = K-2$, then we correct all checks with constraint weight $|\vec{c}| = K-3$ and so on, down to checks with constraint weight $|\vec{c}| = 1$. We finally correct the check $C[\{\vec{0}\}]$ by flipping all bits assuming $C[\{\vec{0}\}]$ is violated. This procedure gives us a valid correction with with all checks satisfied. We go on to find alternative corrections by considering other codewords of our code. In what follows we explain how the correction step proceeds.

We have a set of bits $b[\vec{c} ]$ that describe the results of matching subroutines that may be incorrect. We look for a new set of bits $b'[\vec{v}] = b[\vec{v}] + a[\vec{v}]$ where bits $a[\vec{v}]$ are a correction such that $b'[\vec{v}]$ satisfy all checks $C[\{\vec{c}\}]$. Let us look at how we find an initial solution for $a[\vec{v}]$ before considering alternative solutions. Unless stated otherwise we set all values $a[\vec{v}] = 0$.

We first consider checks $C[\{\vec{c}\}]$ with $|\vec{c} | = K - 2$. If we find a check $C[\{\vec{c}\}]$ is violated, then we set $a[\vec{v} = \vec{c}] = 1$, otherwise we set $a[\vec{v} = \vec{c}] = 0$. We do this for all checks $C[\{ \vec{c} \}]$ with $|\vec{c} | = K - 2$ before continuing to checks with vector sets with fewer constraints. Specifically we continue with checks $C[ \{\vec{c}\} ]$ with incrementally lower constraint weight, i.e., for all checks of the same weight $|\vec{c} |$ in order with $|\vec{c} | = K - 3, \, K-4 , \, \dots, $and so on down to $|\vec{c} | = 1$.

This strategy for finding an initial correction works because, with this ordering, we can change the value of $a[\vec{v} = \vec{c}]$ according to check $C[\{\vec{c}\}]$ without changing the values of any of the checks corrected previously.
In order to see this, we simply check that flipping the bit $a[\vec{v} = \vec{c}]$ will not affect any other checks $C[\{\vec{c} \ ' \}]$ with $|\vec{c} \ '| \ge |\vec{c} |  $. To see this, without loss of generality, we can check that $\vec{v} = \vec{c}$ appears in $\mathcal{V}[\{ \vec{c} \}]$ only, but not in any vector set $\mathcal{V}[\{ \vec{c} \ ' \}]$ with $|\vec{c} \ '| \ge |\vec{c} |$. This is obvious because $\mathcal{V}[\{ \vec{c} \ ' \}]$ contains at most one vector with hamming weight $ |\vec{c} | $. In the case that $ |\vec{c} \ '| = |\vec{c} |  $ the set of vectors  $\mathcal{V}[\{ \vec{c} \ ' \}]$ has only one vector with hamming weight $|\vec{c} |$ which is $\vec{v} = \vec{c} \ '$, which cannot be $\vec{c} \ ' $ since $\vec{c} \ '\not= \vec{c}$. Similarly, in the case that $ |\vec{c} \ '| > |\vec{c} |  $ then clearly the vector set $\mathcal{V}_C[\{ \vec{c} \ ' \}]$ does not contain the vector $\vec{v}  = \vec{c}$ since all of the vectors in this set have hamming weight $ |\vec{c} \ '| > |\vec{c} |  $ and therefore cannot include vector $\vec{v} = \vec{c}$. If follows that we can set $a[\vec{v}=\vec{c}] = 1$ without affecting $C[\{\vec{c} \ ' \} ]$ with constraint weight $| \vec{c} \ ' | \ge | \vec{c}  |$.

Up to this point all we have found a correction such that all checks are satisfied except for $C[\vec{0}]$. In the case that this check is violated, we can correct its value by flipping all bits $a[\vec{v}]$ from the values that have been set up to this point. This will alter the value of the check with constraint set $\vec{c} = \vec{0}$ and no other check since, as we have mentioned, $C[\vec{0}]$ is the only check with an odd support on the bits .
With this final check satisfied, we have a correction that satisfies all checks $C[\{\vec{c}\}]$. It remains to find the best choice of correction. In order to do so, we can toggle between different codewords for the code we have defined.

We toggle between different choices of correction for the code by flipping all correction bits $a[\vec{u}]$ with vectors in the set $\vec{u}\in \mathcal{V}[\{\vec{\ell}\}]$ with a constraint set $\{\vec{\ell}\}$ that satisfies $|\vec{\ell}| = 1$. To check this transforms between codewords, we must check that flipping all bits in this set of vectors will preserve the values of the checks. This problem boils down to checking that the vector set $ \mathcal{V}[\{\vec{\ell}\}]$ has an even number of vectors in common with $ \mathcal{V}[\{\vec{c}\}]$ for check $C[\{\vec{c}\}]$, as this will imply flipping all bits $a[\vec{u}]$ for $ \vec{u}\in \mathcal{V}[\{\vec{\ell}\}]$ and therefore preserve the value of $C[\{\vec{c}\}]$.

We verify this with the following argument. Vectors $\vec{u}\in \mathcal{V}[\{\vec{\ell}\}]$ have exactly one fixed value and all other values are free. Without loss of generality, let us assume that the first element $u_1 = 1$ is fixed for all $\vec{u}\in \mathcal{V}[\{\vec{\ell}\}]$. Otherwise, all vectors for $\vec{u}\in \mathcal{V}[\{\vec{\ell}\}]$ are allowed. We are therefore only interested in bits with $\vec{v}\in \mathcal{V}[\{\vec{c}\}]$ with $v_1 = 1$. This may be because this element is free or fixed for vectors $\vec{v}\in \mathcal{V}[\{\vec{c}\}]$. In either case, it remains for us to look at the possible constraints on elements $v_j$ with $2 \le j \le K$ for vectors $\vec{v}\in \mathcal{V}[\{\vec{c}\}]$. Let us say that $C$ of these remaining $K-1$ elements are fixed to $1$, where $C \le K - 2$. It follows that we have $F := K - 1 - C \ge 1$ free elements for vectors $\vec{v}\in \mathcal{V}[\{\vec{c}\}]$ among elements $v_j$ with $j \ge 2$. As all of these vectors must be included in $ \mathcal{V}[\{\vec{\ell}\}] $, it follows that $\mathcal{V}[\{\vec{c}\}]$ and $\vec{v}\in \mathcal{V}[\{\vec{\ell}\}]$ have $2^F$ terms in common with $F\ge 1$. As this number is even, it follows that $\mathcal{V}[\{\vec{c}\}]$ and $\vec{v}\in \mathcal{V}[\{\vec{\ell}\}]$ share an even number of vectors.

We toggle between different code words of the code by flipping sets of bits with vector indices ins sets $\mathcal{V}[\{\vec{\ell}\}]$ with a constraint sets $\{\vec{\ell}\}$ satisfying $|\vec{\ell}| = 1$. We have $K$ sets $\mathcal{V}[\{\vec{\ell}\}]$, and as such we have $2^K$ combinations of flips we can conduct on the correction bits $a[\vec{v}]$. Once we decide on the best choice of correction bits $a[\vec{v}]$ we can read off the corrected matching results from bits $b'[\vec{\ell}]$ for logical operators $\overline{Z}_j$ where constraint set $|\vec{\ell}| = 1$ is such that $\ell_j = 1$.

\begin{figure}[t]
    \centering
\includegraphics[width=0.99\linewidth]{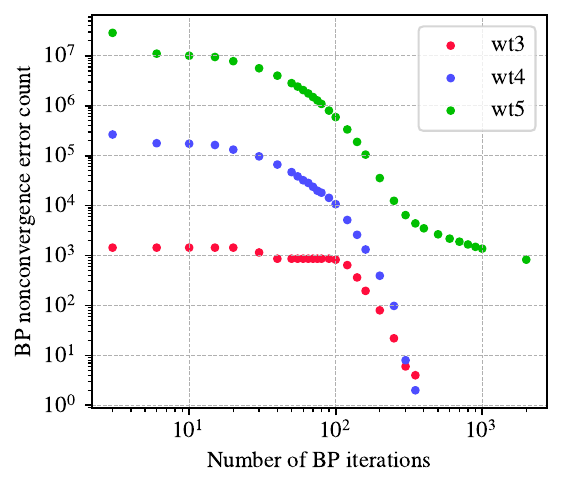}
    \caption{We vary the number of BP iterations on the gross code and exhaustively count the number of low weight errors on which BP fails to converge. 
    }
    \label{fig:bpconvergence}
\end{figure}

\section{Soft information} \label{App:CorrSym}

In this section, we present further analysis on how the use of soft information benefits our decoder.

\begin{figure}[t]
    \centering
\includegraphics[width=0.99\linewidth]{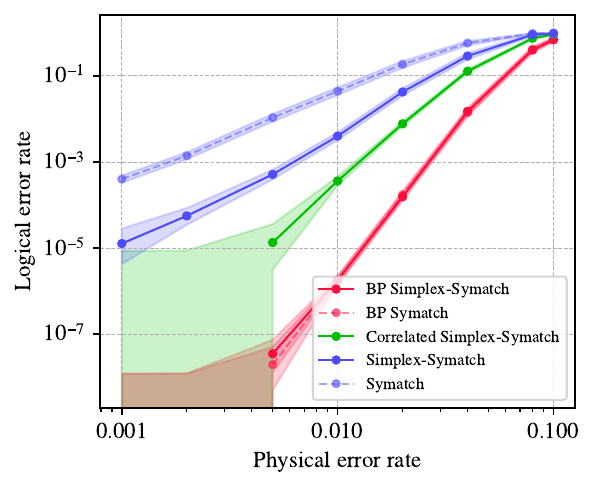}
    \caption{We compare different strategies of message passing between matching subroutines while decoding BB144. The red curves incorporate BP, while the green curve uses correlated matching. 
    }
    \label{fig:corrmatching}
\end{figure}

\subsection{The efficacy of BP}

Belief propagation (BP) utilizes message passing between nodes of the code tanner graph based on the syndrome of a given shot in order to  update the error priors. In cases where the error mechanisms with priors above a certain threshold satisfy the syndrome, we say that BP has converged. 

BP can sometimes converge incorrectly. The convergence rate and accuracy is based on a number of internal choices which include (1) the initial prior assigned to all qubits. (2) the max number of BP iterations allowed, and (3) the BP variant used. Throughout this work,we use the \texttt{min-sum} BP subroutine, and set the \texttt{ms-scaling-factor=0}.  These settings ensure no logically incorrect convergences for low-weight errors. We empirically found a prior of $3/144$ to function well for the exhaustive search in \cref{table:grossexh}.

We can further analyze the number of errors of each weight for which BP does not converge. This data is presented in \cref{fig:bpconvergence}. Clearly, for the hyperparameters stated while decoding BB144, it makes sense to use over 500 iterations of BP, as only weight-5 errors are left for the secondary subroutine to correct.

\subsection{Correlated symatching}

Here, we attempt to pass information between matching subroutines for bit-flip noise in a manner similar to the correlated approach of Refs.~\cite{fowler2013optimalcomplexitycorrectioncorrelated, pymatchingv2}, originally proposed for $Y$ noise. We caution that this is a rudimentary exploration of this idea, and warrants further improvement. Our strategy is as follows: 
\begin{enumerate}
    \item We first perform an initial round of matching on all $2^K$ symmetries. At this stage, for any $\Sigma_i$, all weights in the symmetry graph $\mathcal{G}_{\Sigma_{i}}= (\mathcal{V}_i, \mathcal{E}_i)$ are such that $w(e) = 1 \quad \forall e\in \mathcal{E}_i$
     \item Let symmetry $\Sigma_i$ return an edge subset $\mathcal{E}_i'$ in the initial matching. These edges have a one-to-one correspondence with qubits. Now, in all $\mathcal{G}_{\Sigma_{\neq i}}$, we reduce the weight of these qubits' corresponding edges by $\epsilon$. In other words, in all $\mathcal{G}_{\Sigma_{\neq i}}$
     \begin{equation}
         w(e) \leftarrow w(e) - \epsilon \quad \forall e\in \mathcal{E}'_i  
     \end{equation}
     Intuitively, if a majority of symmetries have a particular qubit in their initial solution, that qubit's corresponding edge wight is maximally suppressed.
     \item We now match again on the reweighted graphs of all $2^K$ symmetries, followed by classical simplex correction of the matching result.
\end{enumerate}

The performance of correlated symatching for the gross code is shown in \cref{fig:corrmatching}. Evidently, there is an improvement over unweighted simplex-symatching. However, it is apparent that BP is able to provide greater overall accuracy. In future work, it would be interesting to explore this design space and investigate different ways of passing matching results across symmetries~\cite{jones2024improvedaccuracydecodingsurface}.

\end{document}